\documentclass[aps,prb,twocolumn,floatfix,showpacs,citeautoscript]{revtex4}

\usepackage{amsmath}
\usepackage{amssymb}
\usepackage{amsfonts}
\usepackage{graphicx,graphics}
\usepackage{bm}

\newcommand{\ket}[1]{\lvert #1\rangle}
\newcommand{\bra}[1]{\langle#1 \rvert}
\newcommand{\abs}[1]{\lvert #1 \rvert}
\newcommand{\expect}[1]{\langle #1\rangle}

\newcommand{\br}{\mathbf{r}}
\newcommand{\bk}{\mathbf{k}}
\newcommand{\bq}{\mathbf{q}}

\begin{document}

\title{First-principles study of the phonon-limited mobility in $n$-type single-layer MoS$_2$}
\author{Kristen Kaasbjerg}
\email{cosby@fys.ku.dk}
\author{Kristian S. Thygesen}
\author{Karsten W. Jacobsen}
\affiliation{Center for Atomic-scale Materials Design (CAMD),\\
  Department of Physics, Technical University of Denmark, 
  DK-2800 Kgs. Lyngby, Denmark}
\date{\today}

\begin{abstract}
  In the present work we calculate the phonon-limited mobility in intrinsic
  $n$-type single-layer MoS$_2$ as a function of carrier density and temperature
  for $T>100$~K. Using a first-principles approach for the calculation of the
  electron-phonon interaction, the deformation potentials and Fr{\"o}hlich
  interaction in the isolated MoS$_2$ layer are determined. We find that the
  calculated room-temperature mobility of $\sim 410$~cm$^2$~V$^{-1}$~s$^{-1}$ is
  dominated by optical phonon scattering via deformation potential couplings and
  the Fr{\"o}hlich interaction with the deformation potentials to the
  intravalley homopolar and intervalley longitudinal optical phonons given by
  $4.1\times10^8$~eV/cm and $2.6\times10^8$~eV/cm, respectively. The mobility is
  weakly dependent on the carrier density and follows a $\mu \sim T^{-\gamma}$
  temperature dependence with $\gamma=1.69$ at room temperature. It is shown
  that a quenching of the characteristic homopolar mode which is likely to occur
  in top-gated samples, boosts the mobility with $\sim
  70$~cm$^2$~V$^{-1}$~s$^{-1}$ and can be observed as a decrease in the exponent
  to $\gamma=1.52$. Our findings indicate that the intrinsic phonon-limited
  mobility is approached in samples where a high-$\kappa$ dielectric that
  effectively screens charge impurities is used as gate oxide.
\end{abstract}

\pacs{81.05.Hd, 72.10.-d, 72.20.-i, 72.80.Jc}
\maketitle

\section{Introduction}

Alongside with the rise of graphene and the exploration of its unique electronic
properties~\cite{Geim:Graphene,RMP:Graphene,Sarma:RMP}, the search for other
two-dimensional (2D) materials with promising electronic properties has gained
increased interest~\cite{Novoselov:RPP}. Metal dichalcogenides which are layered
materials similar to graphite, provide interesting candidates. Their layered
structure with weak interlayer van der Waals bonds allows for fabrication of
single- to few-layer samples using mechanical peeling/cleavage or chemical
exfoliation techniques similar to the fabrication of
graphene~\cite{Geim:2D,Fuhrer:Ultrathin,Pati:Analogues,Kis:MoS2Transistor}.
However, in contrast to graphene they are semiconductors and hence come with a
naturally occurring band gap---a property essential for electronic applications.

The vibrational and optical properties of single- to few-layer samples of
MoS$_2$ have recently been studied extensively using various experimental
techniques~\cite{Ryu:Anomalous,Heinz:ThinMoS2,Schuller:Photocarrier}. In
contrast to the bulk material which has an indirect-gap, it has been
demonstrated that single-layer MoS$_2$ is a direct-gap semiconductor with a gap
of $1.8$ eV~\cite{Heinz:ThinMoS2,Wang:Emerging}. Together with the excellent
electrostatic control inherent of two-dimensional materials, the large band gap
makes it well-suited for low power applications~\cite{Salahuddin:HowGood}. So
far, electrical characterizations of single-layer MoS$_2$ have shown $n$-type
conductivity with room-temperature mobilities in the range $0.5-3$ cm$^2$
V$^{-1}$ s$^{-1}$~\cite{Geim:2D,Fuhrer:Ultrathin,Kis:MoS2Transistor}. Compared
to early studies of the intralayer mobility of bulk MoS$_2$ where mobilities in
the range $100-260$~cm$^2$~V$^{-1}$~s$^{-1}$ were
reported~\cite{Mooser:Mobility}, this is rather low. In a recent experiment the
use of a high-$\kappa$ gate dielectric in a top-gated device was shown to boost
the carrier mobility to a value of
200~cm$^2$~V$^{-1}$~s$^{-1}$~\cite{Kis:MoS2Transistor}. The observed increase in
the mobility was attributed to screening of impurities by the high-$\kappa$
dielectric and/or modifications of MoS$_2$ phonons in the top-gated sample.
\begin{figure}[!b]
  \begin{minipage}{0.46\linewidth}
    \centering
    \includegraphics[width=0.99\linewidth]{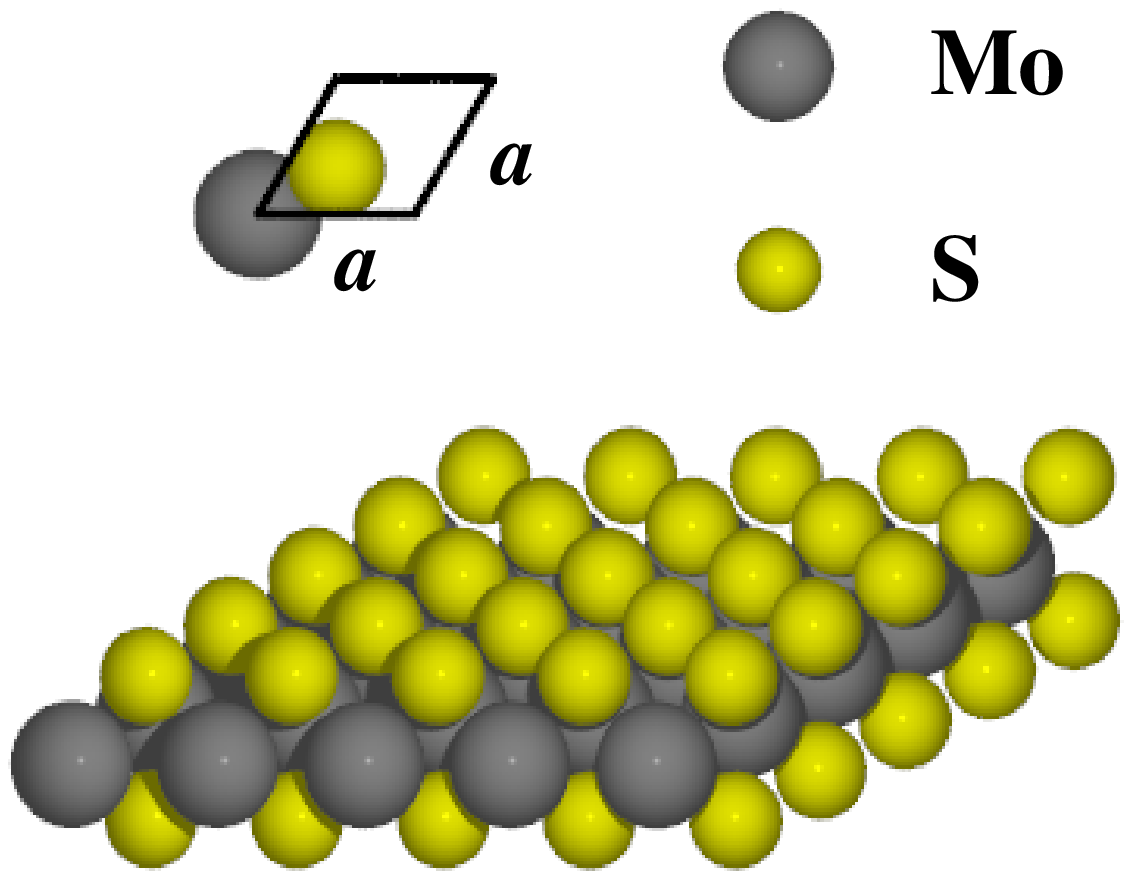}
  \end{minipage} \hfill
  \begin{minipage}{0.52\linewidth}
    \centering
    \includegraphics[width=0.99\linewidth]{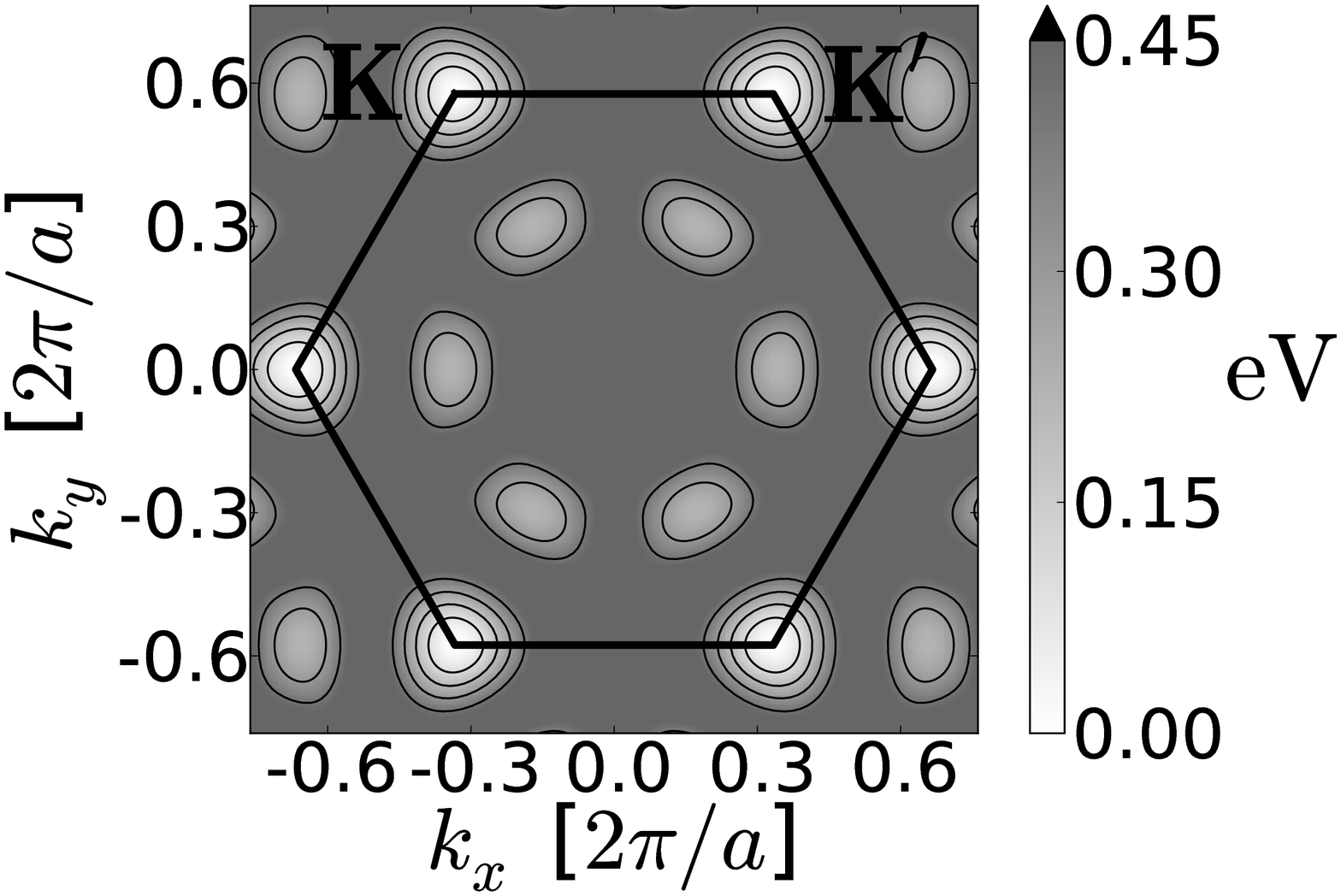}
  \end{minipage} 
  \caption{(Color online) Atomic structure and conduction band of single-layer
    MoS$_2$. Left: Primitive unit cell and structure of an MoS$_2$ layer with
    the molybdenum and sulfur atoms positioned in displaced hexagonal layers.
    Right: Contour plot showing the lowest lying conduction band valleys as
    obtained with DFT-LDA in the hexagonal Brillouin zone of single-layer
    MoS$_2$. For $n$-type MoS$_2$, the low-field mobility is determined by the
    carrier properties in the $K,K'$-valleys which are separated in energy from
    the satellite valleys inside the Brillouin zone.}
\label{fig:mos2}
\end{figure}

In order to shed further light on the measured mobilities and the possible role
of phonon damping due to a top-gate dielectric~\cite{Kis:MoS2Transistor}, we
have carried out a detailed study of the phonon-limited mobility in single-layer
MoS$_2$. Since phonon scattering is an intrinsic scattering mechanism that often
dominates at room temperature, the theoretically predicted phonon-limited
mobility sets an upper limit for the experimentally achievable mobilities in
single-layer MoS$_2$~\cite{Sarma:100}.  In experimental situations, the
temperature dependence of the mobility can be useful for the identification of
dominating scattering mechanisms and also help to establish the value of the
intrinsic electron-phonon couplings~\cite{Sarma:Hetero1}. However, the existence
of additional extrinsic scattering mechanisms often complicates the
interpretation.  For example, in graphene samples impurity scattering and
scattering on surface polar optical phonons of the gate oxide are important
mobility-limiting
factors~\cite{Sarma:2DGraphene,MacDonald:Massless,Fuhrer:GrapheneSiO2,Jena:HighKappa,Kim:SO,Sarma:Nonmonotonic}.
It can therefore be difficult to establish the value and nature of the intrinsic
electron-phonon interaction from experiments alone. Theoretical studies
therefore provide useful information for the interpretation of experimentally
measured mobilities.

In the present work, the electron-phonon interaction in single-layer MoS$_2$ is
calculated from first-principles using a density-functional based approach.
From the calculated electron-phonon couplings, the acoustic (ADP) and optical
deformation potentials (ODP) and the Fr{\"o}hlich interaction in single-layer
MoS$_2$ are inferred. Using these as inputs in the Boltzmann equation, the
phonon-limited mobility is calculated in the high-temperature regime ($T >
100$~K) where phonon scattering often dominate the mobility. Our findings
demonstrate that the calculated phonon-limited room-temperature mobility of
$\sim 410$~cm$^2$~V$^{-1}$~s$^{-1}$ is dominated by optical deformation
potential and polar optical scattering via the Fr{\"o}hlich interaction. The
dominating deformation potentials of $D^0_\text{HP} = 4.1\times10^8$~eV/cm and
$D^0_\text{LO} = 2.6\times10^8$~eV/cm originate from the couplings to the
homopolar and the intervalley polar LO phonon. This is in contrast to previous
studies for bulk MoS$_2$, where the mobility has been assumed to be dominated
entirely by scattering on the homopolar
mode~\cite{Mooser:Mobility,Fivaz:Dimensionality,Schmid:Layered}. Furthermore, we
show that a quenching of the characteristic homopolar mode which is polarized in
the direction normal to the MoS$_2$ layer can be observed as a change in the
exponent $\gamma$ of the generic temperature dependence $\mu \sim T^{-\gamma}$
of the mobility. Such a quenching can be expected to occur in top-gated samples
where the MoS$_2$ layer is sandwiched between a substrate and a gate oxide.

The paper is organized as follows. In Section~\ref{sec:II} the band structure
and phonon dispersion of single-layer MoS$_2$ as obtained with DFT-LDA are
presented. The latter is required for the calculation of the electron-phonon
coupling presented in Section~\ref{sec:III}. From the calculated electron-phonon
couplings, we extract zero- and first-order deformation potentials for both
intra and intervalley phonons. In addition, the coupling constant for the
Fr{\"o}hlich interaction with the polar LO phonon is determined. In
Section~\ref{sec:IV}, the calculation of the phonon-limited mobility within the
Boltzmann equation is outlined. This includes a detailed treatment of the phonon
collision integral from which the scattering rates for the different coupling
mechanisms can be extracted. In Section~\ref{sec:V} we present the calculated
mobilities as a function of carrier density and temperature. Finally, in
Section~\ref{sec:VI} we summarize and discuss our findings.

\section{Electronic structure and phonon dispersion}
\label{sec:II}

The electronic structure and phonon dispersion of single-layer MoS$_2$ have been
calculated within DFT in the LDA approximation using a real-space
projector-augmented wave (PAW) method~\cite{GPAW,GPAW1,GPAW2}. The equilibrium
lattice constant of the hexagonal unit cell shown in Fig.~\ref{fig:mos2} was
found to be $a=3.14$~{\AA} using a $11\times 11$ $\bk$-point sampling of the
Brillouin zone. In order to eliminate interactions between the layers due to
periodic boundary conditions, a large interlayer distance of 10~{\AA} has been
used in the calculations.

\subsection{Band structure} 

In the present study, the band structure of single-layer MoS$_2$ is calculated
within DFT-LDA~\cite{Eriksson:2D,Hong:Interlayer}. While DFT-LDA in general
underestimates band gaps, the resulting dispersion of individual bands, i.e.
effective masses and energy differences between valleys, is less problematic.
We note, however, that recent GW quasi-particle calculations which in general
give a better description of the band structure in standard
semiconductors~\cite{GW}, suggest that the distance between the $K,K'$-valleys
and the satellite valleys located on the $\Gamma$-$K$ path inside the Brillouin
zone not as large as predicted by DFT (see below) and that the valley ordering
may even be inverted~\cite{Ciraci:Functionalization,Olsen:MoS2}. This, however,
seems to contradict with the experimental consensus that single-layer MoS$_2$ is
a direct-gap semiconductor~\cite{Heinz:ThinMoS2,Wang:Emerging} and further
studies are needed to clarify this issue. In the following we therefore use the
DFT-LDA band structure and comment on the possible consequences of a valley
inversion in Section~\ref{sec:VI}.
\begin{figure}[!t]
  \centering
  \includegraphics[width=0.9\linewidth]{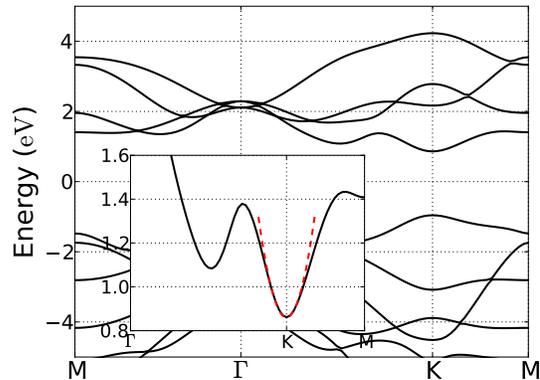}
  \caption{(Color online) Band structure of single-layer MoS$_2$. The inset
    shows a zoom of the conduction band along the path $\Gamma$-K-M. The
    parabolic band (red dashed line) with effective mass $m^*=0.48\;m_e$ is seen
    to give a perfect description of the conduction band valley in the $K$-point
    for the range of energies relevant for the low-field mobility.}
\label{fig:bandstructure}
\end{figure}

The calculated DFT-LDA band structure is shown in Fig.~\ref{fig:bandstructure}.
It predicts a direct band gap of $1.8$~eV at the $K$-poing in the Brillouin
zone. The inset shows a zoom of the bottom of the conduction band along the
$\Gamma$-K-M path in the Brillouin zone. As demonstrated by the dashed line, the
conduction band is perfectly parabolic in the $K$-valley. Furthermore, the
satellite valley positioned on the path between the $\Gamma$-point and the
$K$-point lies on the order of $\sim 200$ meV above the conduction band edge and
is therefore not relevant for the low-field transport.

The right plot in Fig.~\ref{fig:mos2} shows a contour plot of the lowest lying
conduction band valleys in the two-dimensional Brillouin zone. Here, the
$K,K'$-valleys that are populated in $n$-type MoS$_2$, are positioned at the
corners of the hexagonal Brillouin zone. Due to their isotropic and parabolic
nature, the part of the conduction band relevant for the low-field mobility can
to a good approximation be described by simple parabolic bands
\begin{equation}
  \label{eq:parabolic}
  \varepsilon_\bk = \frac{\hbar^2 k^2}{2m^*}
\end{equation}
with an effective electron mass of $m^*= 0.48\;m_e$ and where $k$ is measured
with respect to the $K,K'$-points in the Brillouin zone. The two-dimensional
nature of the carriers is reflected in the constant density of states given by
$\rho_0 = g_s g_v m^* / 2\pi\hbar^2$ where $g_s=2$ and $g_v = 2$ are the spin
and $K,K'$-valley degeneracy, respectively. The large density of states which
follows from the high effective mass of the conduction band in the
$K,K'$-valleys, results in non-degenerate carrier distributions except for very
high carrier densities.

\subsection{Phonon dispersion}

\begin{figure}[!t]
  \includegraphics[width=0.9\linewidth]{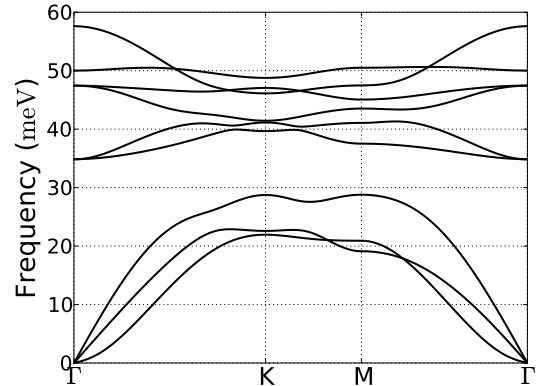}
  \caption{Phonon dispersion of single-layer MoS$_2$ calculated with the small
    displacement method (see e.g. Ref.~\onlinecite{Alfe:SDM}) using a $9\times
    9$ supercell. The frequencies of the two optical Raman active E$_{2g}$ and
    A$_{1g}$ modes at 48 meV and 50 meV, respectively, are in excellent
    agreement with recent experimental measurements~\cite{Ryu:Anomalous}.}
\label{fig:phonon_dispersion}
\end{figure}

The phonon dispersion has been obtained with the supercell-based
small-displacement method~\cite{Alfe:SDM} using a $9\times 9$ supercell. The
resulting phonon dispersion shown in Fig.~\ref{fig:phonon_dispersion} is in
excellent agreement with recent calculations of the lattice dynamics in
two-dimensional MoS$_2$~\cite{Ciraci:Lattice}.

With three atoms in the unit cell, single-layer MoS$_2$ has nine phonon
branches---three acoustic and six optical branches. Of the three acoustic
branches, the frequency of the out-of-plane flexural mode is quadratic in $q$
for $q\rightarrow 0$. In the long-wavelength limit, the frequency of the
remaining transverse acoustic (TA) and longitudinal acoustic (LA) modes are
given by the in-plane sound velocity $c_\lambda$ as
\begin{equation}
  \label{eq:omega_acoustic}
  \omega_{\bq\lambda} = c_\lambda q .
\end{equation}
Here, the sound velocity is found to be $4.4\times 10^3$ and $6.5\times 10^3$
m/s for the TA and LA mode, respectively.

The gap in the phonon dispersion completely separates the acoustic and optical
branches even at the high-symmetry points at the zone-boundary where the
acoustic and optical modes become similar. The two lowest optical branches
belong to the non-polar optical modes. Due to an insignificant coupling to the
charge carriers, they are not relevant for the present study.

The next two branches with a phonon energy of $\sim 48$~meV at the
$\Gamma$-point are the transverse (TO) and longitudinal (LO) polar optical modes
where the Mo and S atoms vibrate in counterphase. In bulk polar materials, the
coupling of the lattice to the macroscopic polarization setup by the lattice
vibration of the polar LO mode results in the so-called LO-TO splitting between
the two modes in the long-wavelength limit. The inclusion of this effect from
first-principles requires knowledge of the Born effective
charges~\cite{Baroni:PhononSemi,Gonze:DynamicalMatrices,Wang:MixedSpace}. In
two-dimensional materials, however, the lack of periodicity in the direction
perpendicular to the layer removes the LO-TO
splitting~\cite{Hernandez:MonolayerBN}. The coupling to the macroscopic
polarization of the LO mode will therefore be neglected here.

The almost dispersionless phonon at $\sim50$ meV is the so-called homopolar mode
which is characteristic for layered structures. The lattice vibration of this
mode corresponds to a change in the layer thickness and has the sulfur layers
vibrating in counterphase in the direction normal to the layer plane while the
Mo layer remains stationary. The change in the potential associated with this
lattice vibration has previously been demonstrated to result in a large
deformation potential in bulk MoS$_2$~\cite{Mooser:Mobility}.

\section{Electron-phonon coupling}
\label{sec:III}

In this section, we present the electron-phonon coupling in single-layer MoS$_2$
obtained with a first-principles based DFT approach. The method is based on a
supercell approach analogous to that used for the calculation of the phonon
dispersion and is outlined in App.~\ref{app:e-ph}. The calculated
electron-phonon couplings are discussed in the context of deformation potential
couplings and the Fr{\"o}hlich interaction well-known from the semiconductor
literature.

Within the adiabatic approximation for the electron-phonon interaction, the
coupling strength for the phonon mode with wave vector $\bq$ and branch index
$\lambda$ is given by
\begin{equation}
  \label{eq:g}
  g_{\bk\bq}^{\lambda} = 
  \sqrt{\frac{\hbar}{2 MN\omega_{\bq\lambda}}} M_{\bk\bq}^\lambda ,
\end{equation}
where $\omega_{\bq\lambda}$ is the phonon frequency, $M$ is an appropriately
defined effective mass, $N$ is the number of unit cells in the crystal, and
\begin{equation}
  \label{eq:M_elph}
  M_{\bk\bq}^\lambda = \bra{\mathbf{k+q}} \delta V_{\bq\lambda}(\br) \ket{\bk}
\end{equation}
is the coupling matrix element where $\bk$ is the wave vector of the carrier
being scattered and $\delta V_{\bq\lambda}$ is the change in the effective
potential per unit displacement along the vibrational normal mode. 

Due to the valley degeneracy in the conduction band, both intra and intervalley
phonon scattering of the carriers in the $K,K'$-valleys need to be considered.
Here, the coupling constants for these scattering processes are approximated by
the electron-phonon coupling at the bottom of the valleys, i.e.  with
$\bk=\mathbf{K},\mathbf{K}'$. With this approach, the intra and intervalley
scattering for the $K,K'$-valleys are thus assumed independent on the wave
vector of the carriers.

In the following two sections, the calculated electron-phonon couplings are
presented~\cite{footnote1}. The different deformation potential couplings are
discussed and the functional form of the Fr{\"o}hlich interaction in 2D
materials is established. Piezoelectric coupling to the acoustic phonons which
occurs in materials without inversion symmetry, is most important at low
temperatures~\cite{Sarma:Hetero2} and will not be considered here. As
deformation potentials are often extracted as empirical parameters from
experimental mobilities, it can be difficult to disentangle contributions from
different phonons. Here, the first-principles calculation of the electron-phonon
coupling allows for a detailed analysis of the couplings and to assign
deformation potentials to the individual intra and intervalley phonons.

\subsection{Deformation potentials}

The deformation potential interaction describes how carriers interact with the
local changes in the crystal potential associated with a lattice vibration.
Within the deformation potential approximation, the electron-phonon coupling is
expressed as~\cite{Ferry}
\begin{equation}
  \label{eq:g_longwavelength}
  g_{\bq\lambda} = \sqrt{\frac{\hbar}{2A\rho\omega_{\bq\lambda}}} M_{\bq\lambda} ,
\end{equation}
where $A$ is the area of the sample, $\rho$ is the atomic mass density per area
and $M_{\bq\lambda}$ is the coupling matrix element for a given valley which is
assumed independent on the $\bk$-vector of the carriers. This expression follows
from the general definition of the electron-phonon coupling in Eq.~\ref{eq:g} by
setting the effective mass $M$ equal to the sum of the atomic masses in the unit
cell. With this convention for the effective mass, $MN=A\rho$, and the
expression in Eq.~\eqref{eq:g_longwavelength} is obtained.

For scattering on acoustic phonons, the coupling matrix element is linear in $q$
in the long-wavelength limit,
\begin{equation}
  \label{eq:M_acoustic}
  M_{\bq\lambda} = \Xi_\lambda q ,
\end{equation}
where $\Xi_\lambda$ is the acoustic deformation potential. 

In the case of optical phonon scattering, both coupling to zero- and first-order
in $q$ must be considered. The interaction via the constant zero-order optical
deformation potential $D_\lambda^0$ is given by
\begin{equation}
  \label{eq:M_optical}
  M_{\bq\lambda} = D_\lambda^0  .
\end{equation}
The coupling via the zero-order deformation potential is dictated by selection
rules for the coupling matrix elements. Therefore, only symmetry-allowed phonons
can couple to the carriers via the zero-order interaction. The coupling via the
first-order interaction is given by Eq.~\eqref{eq:M_acoustic} with the acoustic
deformation potential replaced by the first-order optical deformation potential
$D_\lambda^1$. Both the zero- and first-order deformation potential coupling can
give rise to intra and intervalley scattering.

\begin{figure}[!t]
  \begin{minipage}{1.0\linewidth}
    \vspace{0.25cm} 
    \includegraphics[width=0.25\textwidth]{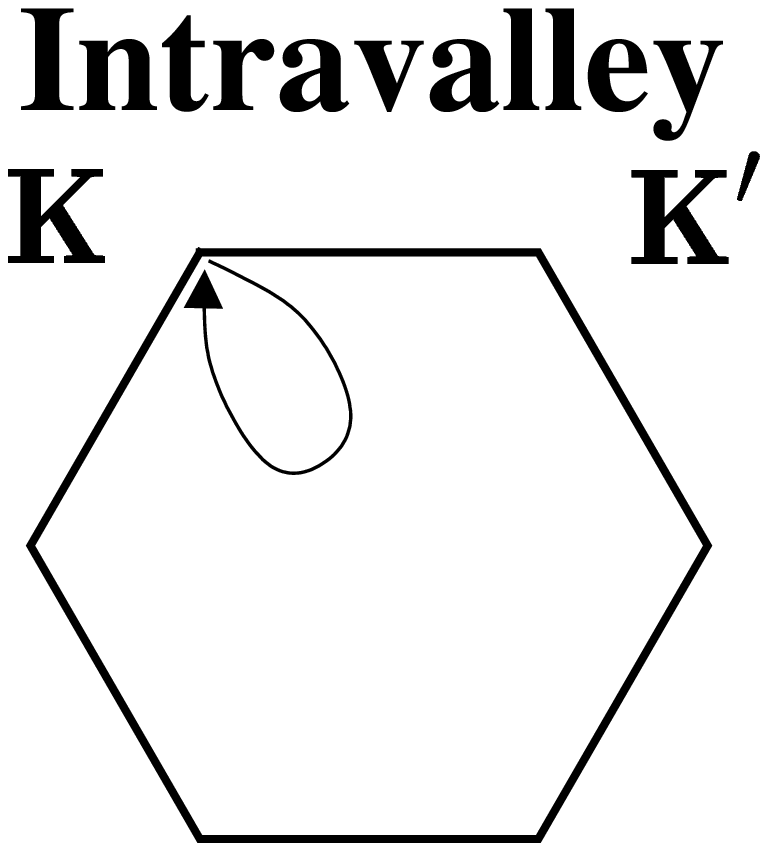}
    \vspace{0.15cm} 
  \end{minipage} 
  \begin{minipage}{1.0\linewidth}
    \includegraphics[width=0.49\textwidth]{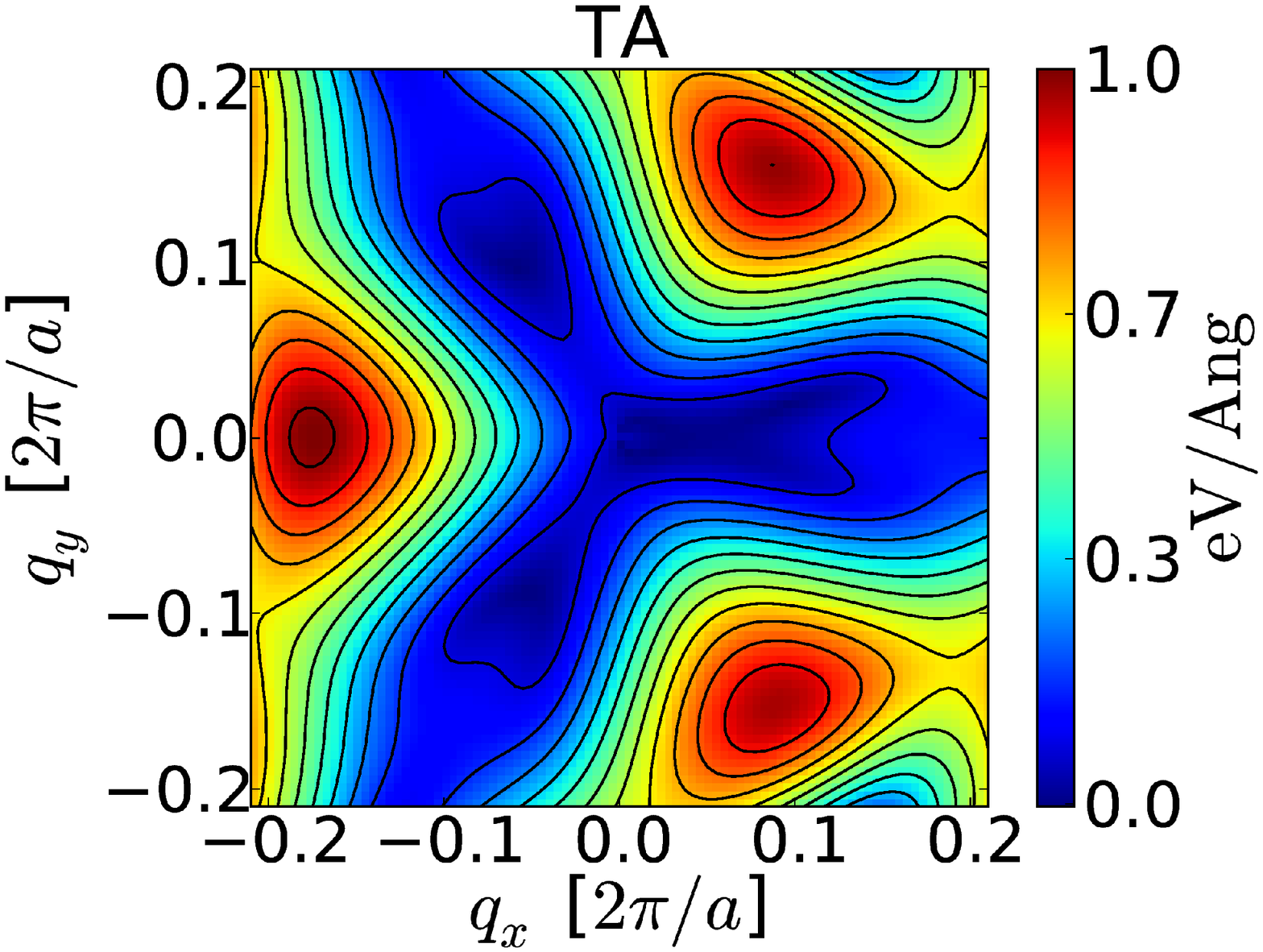}
    \includegraphics[width=0.49\textwidth]{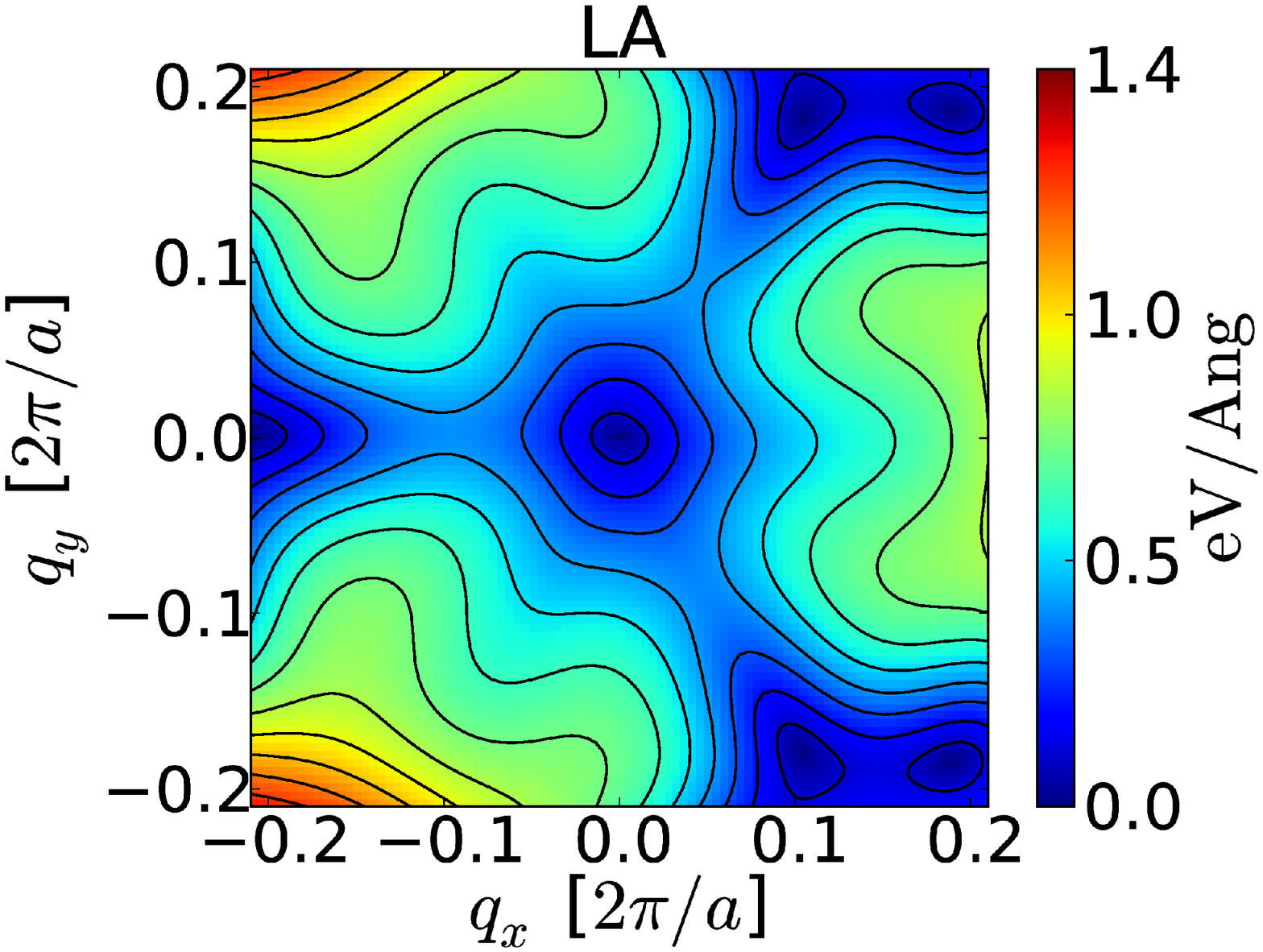}
  \end{minipage}
  \begin{minipage}{1.0\linewidth}
    \includegraphics[width=0.49\textwidth]{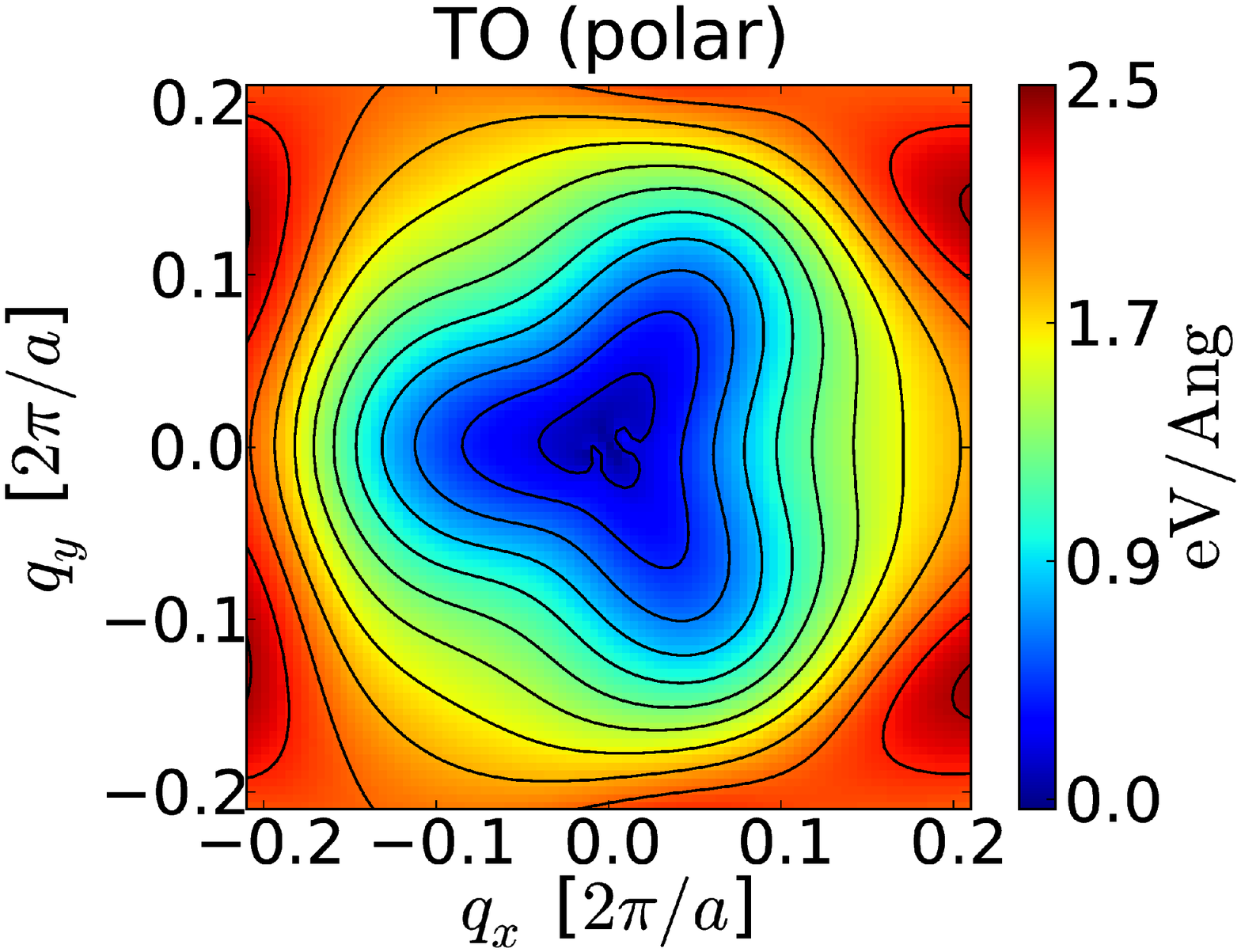}
    \includegraphics[width=0.49\textwidth]{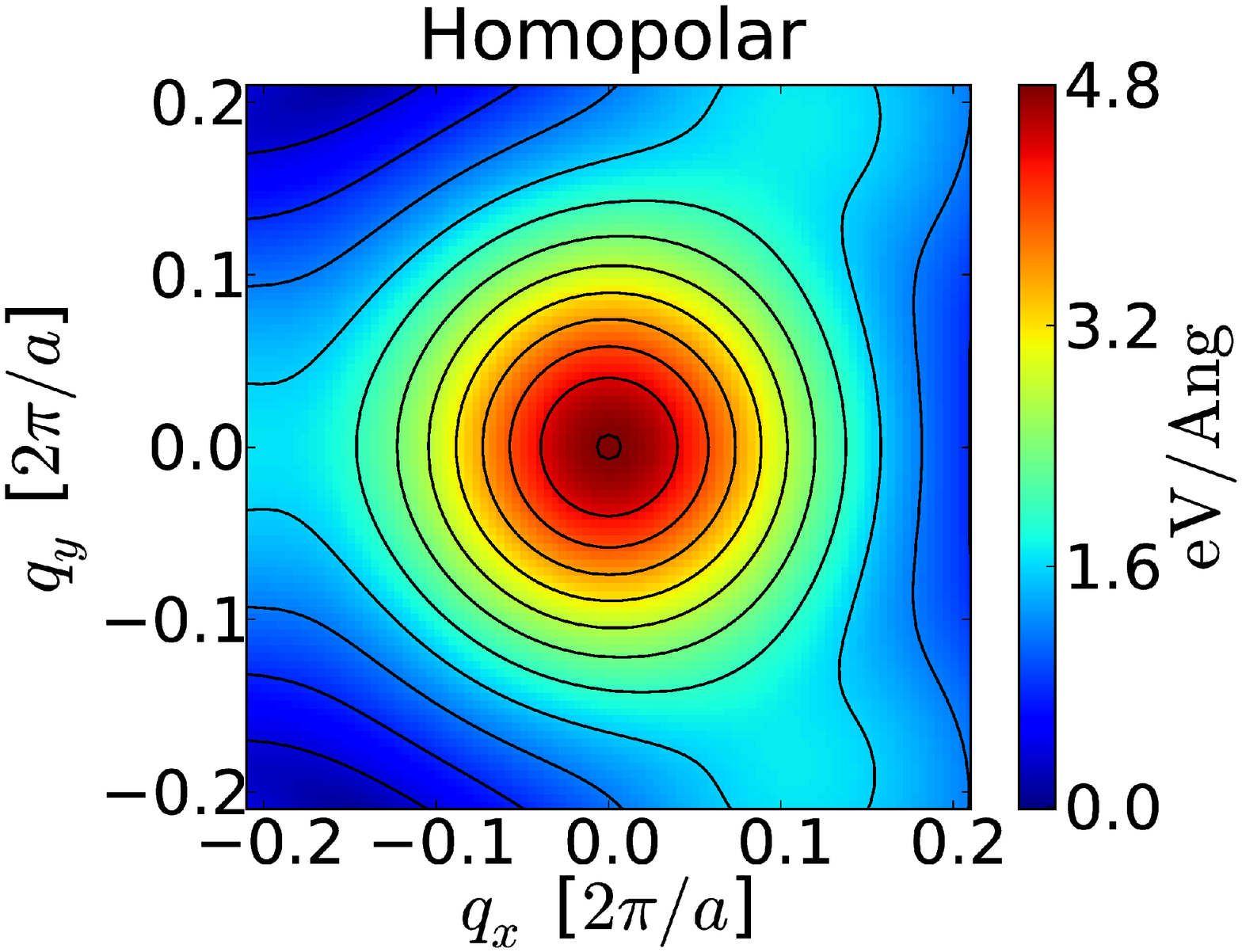}
  \end{minipage} 
  \begin{minipage}{1.0\linewidth}
    \vspace{0.25cm} 
    \includegraphics[width=0.25\textwidth]{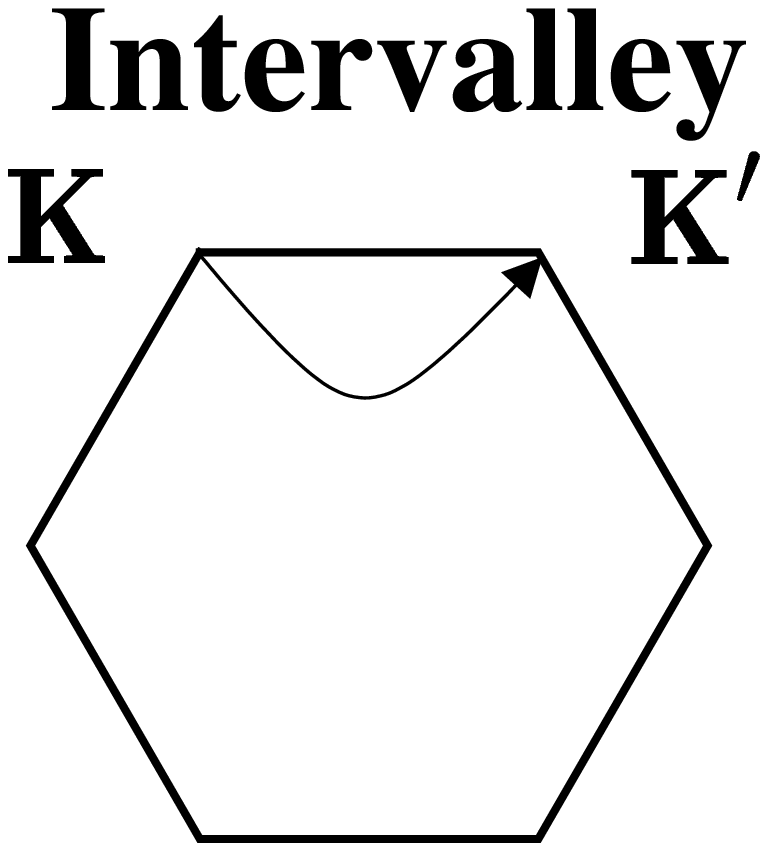}
    \vspace{0.15cm} 
  \end{minipage} 
  \begin{minipage}{1.0\linewidth}
    \includegraphics[width=0.49\textwidth]{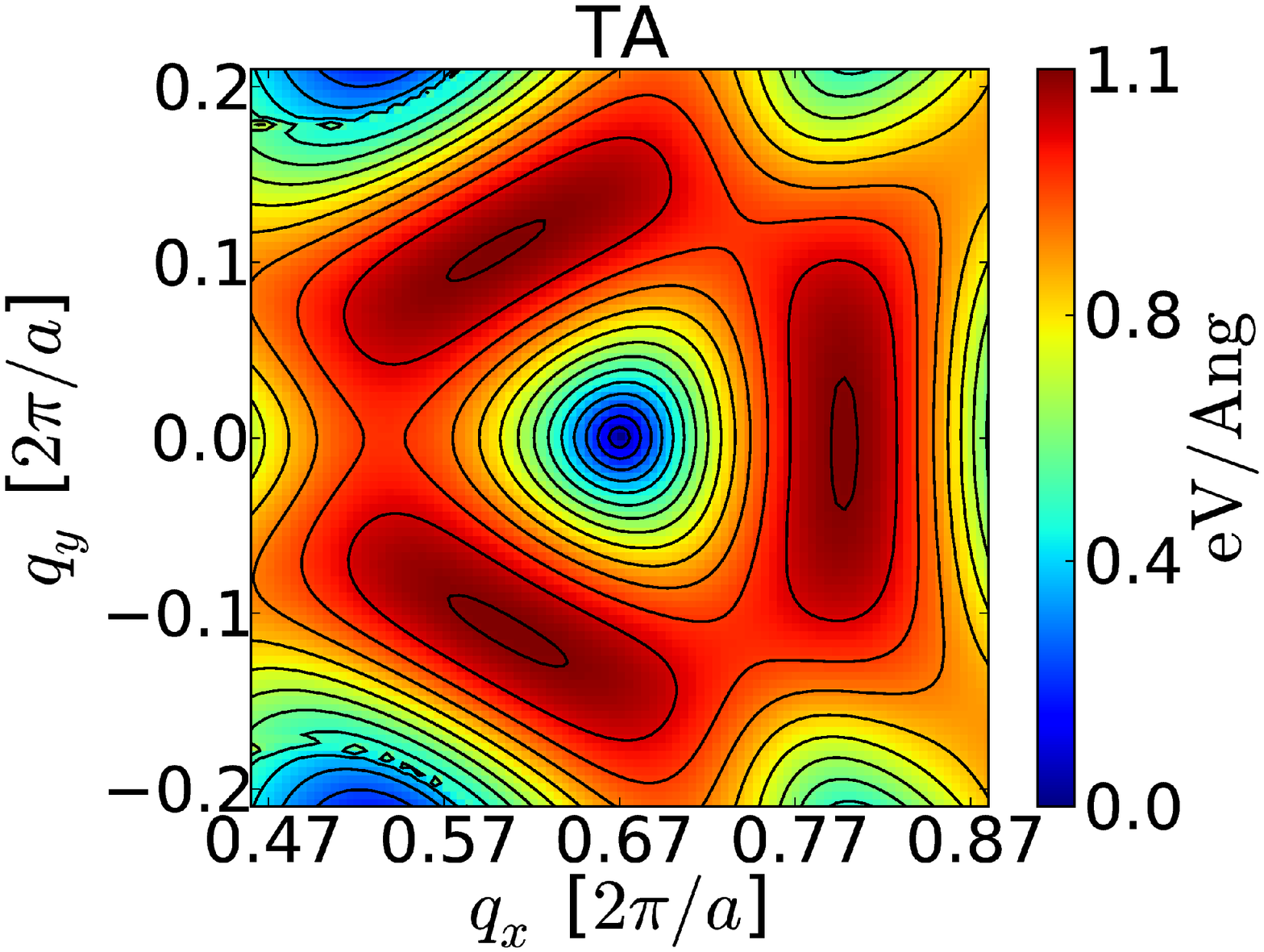}
    \includegraphics[width=0.49\textwidth]{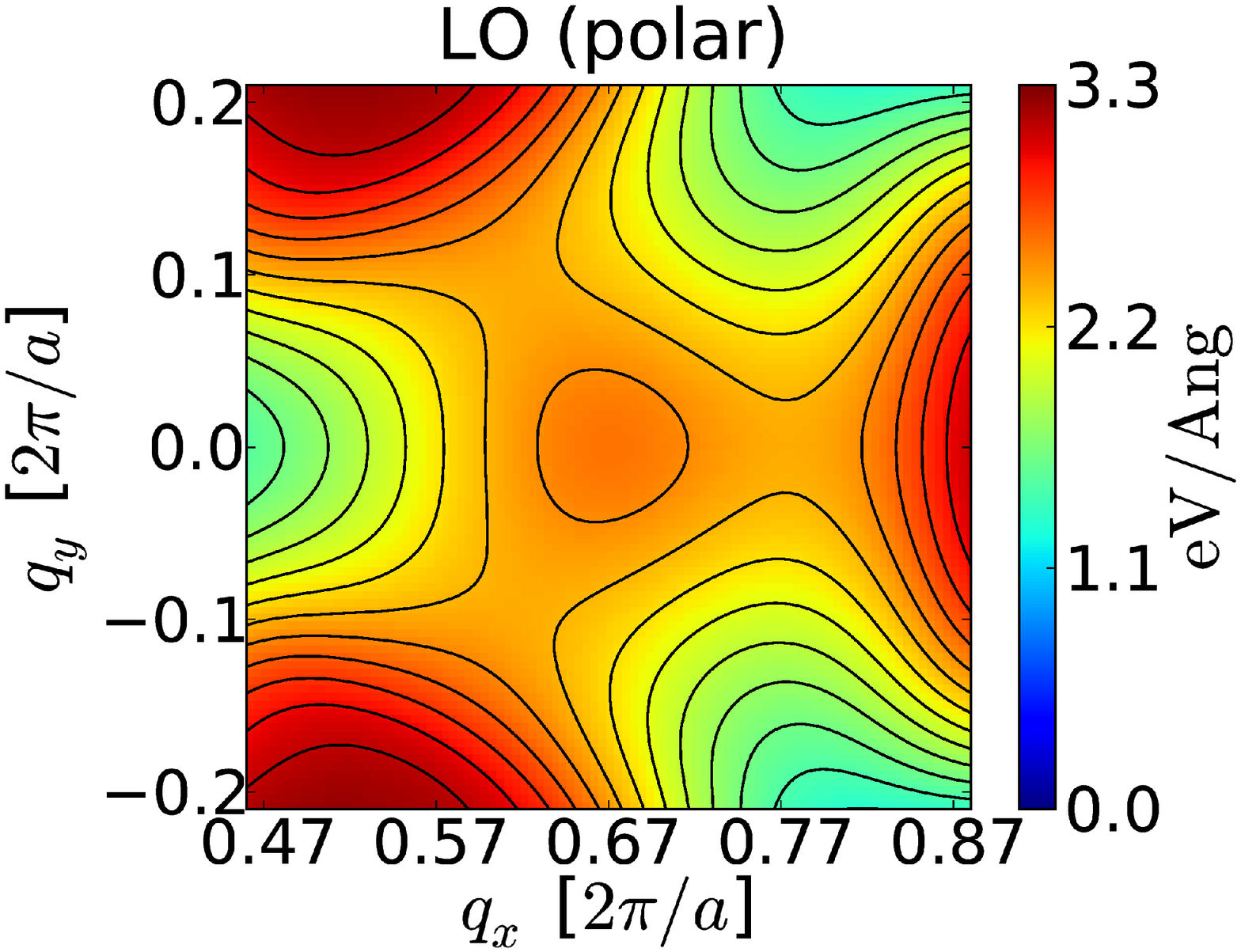}
  \end{minipage}
  \caption{(Color online) Deformation potential couplings in single-layer
    MoS$_2$. Only phonon modes with significant coupling strength are shown.
    The contour plots show the absolute value of the coupling matrix elements
    $\abs{M_{\bq\lambda}}$ in Eq.~\eqref{eq:M_elph} with $\bk=\mathbf{K}$ and
    $\bk' = \mathbf{K} + \bq$ as a function of the two-dimensional phonon wave
    vector $\bq$ (note the different color scales in the plots). The upper four
    plots correspond to intravalley scattering and the two bottom plots to
    intervalley scattering on phonons with wave vector $\bq = 2\mathbf{K}$. The
    two zero-order deformation potential couplings to the intravalley homopolar
    and intervalley LO phonon play an important role for the room-temperature
    mobility.}
\label{fig:couplings}
\end{figure}
The absolute value of the calculated coupling matrix elements
$\abs{M_{\bq\lambda}}$ are shown in Fig.~\ref{fig:couplings} for phonons which
couple to the carriers via deformation potential interactions. They are plotted
in the range of phonon wave vectors relevant for intra and intervalley
scattering processes~\cite{footnote2}, and only phonon modes with significant
coupling strengths have been included. Although the coupling matrix elements are
shown only for $\bk = \mathbf{K}$, the matrix elements for $\bk = \mathbf{K'}$
are related through time-reversal symmetry as
$M_{\bq\lambda}^\mathbf{K}=M_{-\bq\lambda}^\mathbf{K'}$. The three-fold
rotational symmetry of the coupling matrix elements in $\bq$-space stems from
the symmetry of the conduction band in the vicinity of the $K,K'$-valleys (see
Fig.~\ref{fig:mos2}). Since the symmetry of the matrix elements has not been
imposed by hand, slight deviations from the three-fold rotational symmetry can
be observed.

The acoustic deformation potential couplings for the TA and LA modes are shown
in the two top plots of Fig.~\ref{fig:couplings}. Due to the inclusion of
Umklapp processes here, the coupling to the TA mode does not vanish as is most
often assumed~\cite{Madelung}. Only along high-symmetry directions in the
Brillouin zone is this the case. This results in a highly anisotropic coupling
to the TA mode. On the other hand, the deformation potential coupling for the LA
mode is perfectly isotropic in the long-wavelength limit but also becomes
anisotropic at shorter wavelengths. In agreement with Eq.~\eqref{eq:M_acoustic},
both the TA and LA coupling matrix elements are linear in $q$ in the
long-wavelength limit.

The next two plots show the couplings to the intravalley polar TO and homopolar
optical modes. While the interaction with the polar TO phonon corresponds to a
first-order optical deformation potential, the interaction with the homopolar
mode acquires a finite value of $\sim 4.8$~eV/{\AA} in the $\Gamma$-point and
corresponds to a strong zero-order deformation potential coupling. The large
deformation potential coupling to the homopolar mode stems from its
characteristic lattice vibration polarized in the direction perpendicular to the
layer corresponding to a change in the layer thickness. The associated change in
the effective potential from the counterphase oscillation of the two negatively
charged sulfur layers results in a significant change of the potential towards
the center of the MoS$_2$ layer. As the electronic Bloch functions have
significant weight here, the homopolar lattice vibration gives rise to a
significant shift of the $K,K'$-valley states. The large deformation potential
associated with the coupling to the homopolar mode is also present in bulk
MoS$_2$~\cite{Mooser:Mobility}.

The two bottom plots in Fig.~\ref{fig:couplings} show the couplings for the
intervalley TA and polar LO phonons. Due to the nearly constant frequency of the
intervalley acoustic phonons, their couplings are classified as optical
deformation potentials in the following. Both the shown coupling to the
intervalley TA phonon and the coupling to the intervalley LA phonon result in
first-order deformation potentials. The intervalley coupling for the polar LO
phonon gives rise to a zero-order deformation potential which results from an
identical in-plane motion of the two sulfur layers in the lattice vibration of
the intervalley phonon.

In general, the deformation potential approximation, i.e. the assumption of
isotropic and constant/linear coupling matrix elements, in
Eqs.~\eqref{eq:M_acoustic} and~\eqref{eq:M_optical} is seen to hold in the
long-wavelength limit only. At shorter wavelengths, the first-principles
couplings become anisotropic and have a more complicated $q$-dependence. When
determined experimentally from e.g. the temperature dependence of the
mobility~\cite{Sarma:Hetero1}, the deformation potentials in
Eqs.~\eqref{eq:M_acoustic} and~\eqref{eq:M_optical} implicitly account for the
more complex $\bq$-dependence of the \emph{true} coupling matrix element. In
Section~\ref{sec:V}, the theoretical deformation potentials for single-layer
MoS$_2$ are determined from the first-principles electron-phonon couplings. This
is done by fitting their associated scattering rates to the scattering rates
obtained with the first-principles coupling matrix elements. The resulting the
deformation potentials, can be used in practical transport calculations based on
e.g. the Boltzmann equation or Monte Carlo simulations. It should be noted that
similar routes for first-principles calculations of deformation potentials have
been given in the
literature~\cite{Sjakste:GaAsInter,Sjakste:Silicon,Kim:ElPhGraphene}.

\subsection{Fr{\"o}hlich interaction}

The lattice vibration of the polar LO phonon gives rise to a macroscopic
electric field that couples to the charge carriers. For bulk three-dimensional
systems, the coupling to the field is given by the Fr{\"o}hlich interaction
which diverges as $1/q$ in the long-wavelength limit~\cite{Madelung},
\begin{equation}
  \label{eq:frohlich}
  g_\text{LO}(q) = \frac{1}{q}
                   \sqrt{\frac{e^2 \hbar\omega_\text{LO}}{2\epsilon_0 V }}
                   \left(
                     \frac{1}{\varepsilon^\infty} - 
                     \frac{1}{\varepsilon^0}
                   \right)^{1/2} ,
\end{equation}
where $\epsilon_0$ is the vacuum permittivity, $V$ is the volume of the sample
, and $\varepsilon^\infty$ and $\varepsilon^0$ are the high-frequency optical and
static dielectric constant, respectively.

In atomically thin materials, the two-dimensional nature of both the LO phonon
and the charge carriers leads to a qualitatively different $q$-dependence of the
Fr{\"o}hlich interaction. The situation is similar to 2D semiconductor
heterostructures where the Fr{\"o}hlich interaction has been studied using
dielectric continuum models and microscopic based
approaches~\cite{Ando:ElphHetero,Lugli:Microscopic}. From the microscopic
considerations presented in App.~\ref{app:frohlich}, we derive the following
functional form for the Fr{\"o}hlich interaction to the polar LO phonon in 2D
materials,
\begin{equation}
  \label{eq:g_Fr}
  g_\text{LO}(q) = g_\text{Fr} \times \text{erfc}(q \sigma / 2) .
\end{equation}
Here, the coupling constant $g_\text{Fr}$ is the equivalent of the square root
factors in Eq.~\eqref{eq:frohlich}, $\sigma$ is the effective width of the
electronic Bloch states and $\text{erfc}$ is the complementary error function.

\begin{figure}[!t]
  \includegraphics[width=0.8\linewidth]{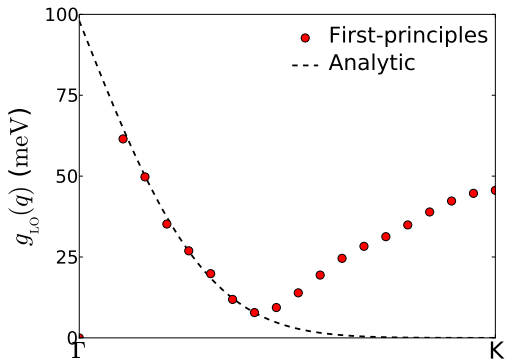}
  \caption{(Color online) Fr{\"o}hlich interaction for the polar optical LO
    mode. The dashed line shows the analytical coupling in Eq.~\eqref{eq:g_Fr}
    with the coupling constant $g_\text{Fr} = 98$~meV and the effective layer
    thickness $\sigma = 4.41$~{\AA} fitted to the long-wavelength limit of the
    calculated coupling (dots).}
\label{fig:frohlich}
\end{figure}

Figure~\ref{fig:frohlich} shows the first-principles coupling $g_\text{LO}$ to
the polar LO phonon in single-layer MoS$_2$ (dots) as a function of the phonon
wave vector along the $\Gamma$-$K$ path. The dashed line shows a fit of the
coupling in Eq.~\eqref{eq:g_Fr} to the long-wavelength limit of the calculated
coupling. With a coupling constant of $g_\text{Fr} = 98$~meV and an effective
width of $\sigma=4.41$~{\AA}, the analytic form gives a perfect description of
the calculated coupling in the long-wavelength limit. For shorter wavelengths,
the zero-order deformation potential coupling to the intervalley LO phonon from
Fig.~\ref{fig:couplings} becomes dominant, and a deviation away from the
behavior in Eq.~\eqref{eq:g_Fr} is observed.

\section{Boltzmann equation}
\label{sec:IV}

In the regime of diffuse transport dominated by phonon scattering, the mobility
can be obtained with semiclassical Boltzmann transport theory. In the absence of
spatial gradients the Boltzmann equation for the out-of-equilibrium distribution
function $f_\bk$ of the charge carriers reads~\cite{SmithJensen}
\begin{equation}
  \label{eq:boltzmann}
  \frac{\partial f_\bk}{\partial t} +
  \dot{\bk} \cdot \nabla_\bk f_\bk
  =
  \left. \frac{\partial f_\bk}{\partial t} \right\vert_{\text{coll}} ,
\end{equation}
where the time evolution of the crystal momentum is governed by $\hbar\dot{\bk}
= q \mathbf{E}$ and $q$ is the charge of the carriers. In the case of a
time-independent uniform electric field, the steady-state version of the
linearized Boltzmann equation takes the form
\begin{align}
  \label{eq:linearized}
  \frac{q}{\hbar} \mathbf{E} \cdot \nabla_{\bk} f_\bk^0
  & = - \frac{q}{k_\text{B}T} \mathbf{E} \cdot \mathbf{v}_\bk f_\bk^0(1 - f_\bk^0)
      \nonumber \\
  & = \left. \frac{\partial f_\bk}{\partial t} \right\vert_{\text{coll}} ,
\end{align}
where $f_\bk^0=f^0(\varepsilon_\bk)$ is the equilibrium Fermi-Dirac distribution
function, $\partial f_\bk^0/\partial \varepsilon_\bk = - f_\bk^0(1 -
f_\bk^0)/k_\text{B}T$, and $\mathbf{v}_\bk = \nabla_{\bk} \varepsilon_\bk /
\hbar$ is the band velocity. In linear response, the out-of-equilibrium
distribution function can be written as the equilibrium function plus a small
deviation $\delta f$ away from its equilibrium value~\cite{SmithJensen}, i.e.
\begin{equation}
  \label{eq:deviation}
  \delta f_\bk = f_\bk - f_\bk^0 = f_\bk^0 (1 - f_\bk^0 ) \psi_\bk .
\end{equation}
Here, the last equality has defined the deviation function $\psi_\bk$.
Furthermore, following the spirit of the iterative method of Rode~\cite{Rode},
the angular dependence of the deviation function is separated out as
\begin{equation}
  \label{eq:Rode}
  \psi_\bk = \phi_k \cos\theta_\bk  ,
\end{equation}
where $\theta_\bk$ is the angle between the $\bk$-vector and the force exerted
on the carriers by the applied field $\mathbf{E}$. In general, $\phi_\bk$ is a
function of the $\bk$-vector and not only its magnitude $k$. However, for
isotropic bands as considered here, the angular dependence of the deviation
function is entirely accounted for by the cosine factor in
Eq.~\eqref{eq:Rode}~\cite{Ferry}.

Considering the phonon scattering, the collision integral describes how
accelerated carriers are driven back towards their equilibrium distribution by
scattering on acoustic and optical phonons. In the high-temperature regime of
interest here, the collision integral can be split up in two contributions. The
first, accounting for the quasielastic scattering on acoustic phonons, can be
expressed in the form of a relaxation time $\tau_{\text{el}}$. The second,
describing the inelastic scattering on optical phonons, will in general be an
integral operator $I_{\text{inel}}$. The collision integral can thus be written
\begin{equation}
  \label{eq:collision}
  \left. \frac{\partial f_\bk}{\partial t} \right\vert_{\text{coll}}
  = - \frac{\delta f_\bk} {\tau_\bk^\text{el}} + I_{\text{inel}} \left[ \psi_\bk \right] ,
\end{equation}
where the explicit forms of the relaxation time and the integral operator will
be considered in detail below.

With the above form of the collision integral, the Boltzmann equation can be
solved by iterating the equation
\begin{align}
  \label{eq:Rode_solution}
  \phi_k = & \left[ \frac{eE v_\bk}{k_\text{B}T} +
                    I'_{\text{inel}} \left[ \phi_k \right]
             \right]
    \times \left( \frac{1}{\tau_\bk^{\text{el}}} \right)^{-1} 
\end{align}
for the deviation function $\phi_k$. Here, the angular dependence
$\cos\theta_\bk$ and a $f_\bk^0 (1 - f_\bk^0 )$ factor have been divided out and
the new integral operator
$I'_{\text{inel}}$ is defined by
\begin{equation}
  \label{eq:I'}
  I_{\text{inel}} = f_\bk^0 (1 - f_\bk^0 ) \cos\theta_\bk I'_{\text{inel}} .
\end{equation}

From the solution of the Boltzmann equation the current and drift mobility of
the carriers can be obtained. Taking the electric field to be oriented along the
$x$-direction, the current density is given by
\begin{equation}
  \label{eq:current_density}
  j_x \equiv \sigma_{xx} E_x  
      =  -\frac{4 e}{A} \sum_\bk v_{x} \delta f_\bk ,
\end{equation}
where $\sigma_{xx}$ is the conductivity, $A$ is the area of the sample, and
$v_{i} = \hbar k_i/m^*$ is the band velocity for parabolic bands with $i= x, y,
z$. From the definition $\mu_{xx} = \sigma_{xx} / ne$ of the drift mobility, it
follows that
\begin{equation}
  \label{eq:mobility}
  \mu_{xx} = \frac{e \expect{\tilde{\phi}_k}}{m^*} 
\end{equation}
where the modified deviation function $\tilde{\phi}_k$ having units of time is
defined by
\begin{equation}
  \label{eq:phi_tilde}
  \tilde{\phi}_k = \frac{m^* k_\text{B}T}{e E_x \hbar k} \phi_k   ,
\end{equation}
and the energy-weighted average $\expect{\cdot}$ is defined by
\begin{equation}
  \label{eq:tau_averaged}
  \expect{A} = \frac{1}{n} 
      \int \! d\varepsilon_\bk \; \rho(\varepsilon_\bk) \varepsilon_\bk
      A_\bk 
      \left(- \frac{\partial f^0}{\partial \varepsilon_\bk} \right) .
\end{equation}
Here, $n$ is the two-dimensional carrier density. In the relaxation time
approximation $\tilde{\phi}_k = \tau_k$ and the well-known Drude expression for
the mobility $\mu = e\expect{\tau_k}/m^*$ is recovered.

\subsection{Phonon collision integral}

The phonon collision integral has been considered in great detail in the
literature for two-dimensional electron gases in semiconductor heterostructure
(see e.g. Ref.~\onlinecite{Sarma:Hetero2}). In general, these treatments
consider scattering on three-dimensional or quasi two-dimensional phonons. In
atomically thin materials the phonons are strictly two-dimensional which results
in a slightly different treatment. In the following, a full account of the
two-dimensional phonon collision integral is given for scattering on acoustic
and optical phonons.

With the distribution function written on the form in Eq.~\eqref{eq:deviation},
the linearized collision integral for electron-phonon scattering takes the
form~\cite{SmithJensen}
\begin{widetext}
\begin{align}
  \label{eq:collision_linearized}
  \left. \frac{\partial f_\bk}{\partial t} \right\vert_{\text{coll}}
  = -\frac{2\pi}{\hbar} \sum_{\bq\lambda} 
    \frac{\left\vert g_{\bq\lambda} \right\vert^2} {\epsilon^2(q, T)}  \times
  &  \bigg[ f_\bk^0 \left( 1 - f_{\bk + \bq}^0 \right)
       N_{\bq\lambda}^0 \left( \psi_\bk - \psi_{\bk + \bq} \right)
        \delta(\varepsilon_{\bk + \bq} - \varepsilon_\bk -
        \hbar\omega_{\bq\lambda})
       \nonumber \\
  & + f_\bk^0 \left( 1 - f_{\bk - \bq}^0 \right)
        \left(1 + N_{\bq\lambda}^0 \right) \left( \psi_\bk - \psi_{\bk - \bq} \right)
        \delta(\varepsilon_{\bk - \bq} - \varepsilon_\bk +
        \hbar\omega_{\bq\lambda})
      \bigg] 
\end{align}
\end{widetext}
where $N_{\bq\lambda}^0=N^0(\hbar\omega_{\bq\lambda})$ is the equilibrium
distribution of the phonons given by the Bose-Einstein distribution function and
$f_{\bk \pm \bq}^0 = f^0(\varepsilon_{\bk} \pm \hbar\omega_{\bq})$ is
understood. The different terms inside the square brackets account for
scattering out of ($\psi_\bk$) and into ($\psi_{\bk\pm\bq}$) the state with wave
vector $\bk$ via absorption and emission of phonons. Screening of the
electron-phonon interaction by the carriers themselves is accounted for by the
static dielectric function $\epsilon$, which is both wave vector $q$ and
temperature $T$ dependent. Due to the large effective mass of the conduction
band, the charge carriers in single-layer MoS$_2$ are non-degenerate except at
very high carrier concentrations. Semiclassical screening where the screening
length is given by the inverse of the Debye-H{\"u}ckel wave vector $k_\text{D} =
e^2 n / 2k_\text{B}T$ therefore applies~\cite{FerryGoodnick}. For the considered
values of carrier densities and temperatures, this corresponds to a small
fraction of the size of the Brillouin zone, i.e. $k_\text{D} \ll 2\pi / a$. As
scattering on phonons in general involves larger wave vectors, screening by the
carriers can to a good approximation be neglected here. The calculated
mobilities therefore provide a lower limit for the proper screened mobility.

\subsubsection{Quasielastic scattering on acoustic phonons} 

For quasielastic scattering on acoustic phonons at high temperatures, the energy
of the acoustic phonon can be neglected in the collision integral in
Eq.~\eqref{eq:collision_linearized}. As a consequence, the collision integral
can be recast in the form of the following relaxation time~\cite{Sarma:Hetero2}
\begin{equation}
  \label{eq:tau_elastic}
  \frac{1}{\tau_\bk^{\text{el}}} = 
     \sum_{\bk'} 
    \left( 1 - \cos{\theta_{\bk\bk'}} \right) P_{\bk\bk'}
\end{equation}
where the summation variable has been changed to $\bk'=\bk\pm \bq$ and the
transition matrix element for quasielastic scattering on acoustic phonons is
given by
\begin{align}
  \label{eq:P_acoustic}
  P_{\bk\bk'} = \frac{2\pi}{\hbar} \sum_\bq
    \left\vert g_{\bq\lambda} \right\vert^2 
    \bigg[ &
    N_{\bq\lambda}^0 
    \delta(\varepsilon_{\bk'} - \varepsilon_\bk) \bigg. \nonumber \\
    & + \bigg. \left(1 + N_{\bq\lambda}^0 \right) 
    \delta(\varepsilon_{\bk'} - \varepsilon_{\bk})
    \bigg]  .
\end{align}
For isotropic scattering as assumed here in Eq.~\eqref{eq:M_acoustic}, the
square of the electron-phonon coupling can be expressed as
\begin{equation}
  \label{eq:g2_elastic}
  \left\vert g_{\bq\lambda} \right\vert^2 = \frac{\Xi_\lambda^2\hbar q}{2A\rho c_\lambda}, 
\end{equation}
where the acoustic phonon frequency has been expressed in terms of the sound
velocity $c_\lambda$. Except at very low temperatures $\hbar\omega_\bq \ll
k_\text{B}T$ implying that the equipartition approximation $N^0_\bq \sim
k_\text{B}T / \hbar\omega_\bq \gg 1$ for the Bose-Einstein distribution applies.
With the resulting $\bq$-factors in the transition matrix element in
Eq.~\eqref{eq:P_acoustic} canceling, the $\cos{\theta_{\bk\bk'}}$ in the
$\bk'$-sum in Eq.~\eqref{eq:tau_elastic} vanishes and the first term yields a
factor density of states divided by the spin and valley degeneracies
$\rho_0/g_sg_v$. The relaxation time for acoustic phonon scattering becomes
\begin{equation}
  \label{eq:tau_acoustic}
  \frac{1}{\tau_\bk^\text{el}} 
  = \frac{m^*\Xi^2k_\text{B}T}{\hbar^3\rho c_\lambda^2} .
\end{equation}
The independence on the carrier energy and the $\tau^{-1} \sim T$ temperature
dependence of the acoustic scattering rate are characteristic for charge
carriers in two-dimensional heterostructures and layered
materials~\cite{Mooser:Mobility,Sarma:Hetero2}.

\subsubsection{Inelastic scattering on dispersionless optical phonons}

For inelastic scattering on optical phonons the phonon energy can no longer be
neglected and the collision integral in Eq.~\eqref{eq:collision_linearized} must
be considered in full detail. Under the reasonable assumption of dispersionless
optical phonons $\omega_{\bq\lambda}=\omega_\lambda$, the inelastic collision
integral can be treated semianalytically. The overall procedure for the
evaluation of the collision integral is given below, while the calculational
details have been collected in App.~\ref{app:collision}. The resulting
expressions for the collision integral apply to both intravalley and intervalley
optical and intervalley acoustic phonons.

In the following, the integral operator for inelastic scattering in
Eq.~\eqref{eq:I'} is split up in separate out- and in-scattering contributions,
\begin{equation}
  \label{eq:I_inel_out_in}
  {I'}_\text{inel}[\phi_k] 
  = {I'}_\text{inel}^\text{out}[\phi_k] + 
    {I'}_\text{inel}^\text{in}[\phi_k]  ,
\end{equation}
which include the terms in Eq.~\eqref{eq:collision_linearized} involving
$\psi_\bk$ and $\psi_{\bk \pm \bq}$, respectively. With the contributions from
different phonon branches $\lambda$ adding up, scattering on a single phonon
with branch index $\lambda$ is considered in the following.

From Eq.~\eqref{eq:collision_linearized}, the out-scattering part of the
collision integral follows directly as
\begin{equation}
  \label{eq:I_out}
  {I'}_\text{inel}^\text{out}[\phi_k] = 
      - \phi_k \sum_{\bk'} P_{\bk\bk'} 
                 \frac{1 - f_{\bk'}^0}{1 - f_\bk^0} ,
\end{equation}
where the transition matrix element for optical phonon scattering is given by
\begin{align}
  \label{eq:P_optical}
  P_{\bk\bk'} = \frac{2\pi}{\hbar}  \sum_\bq
    \left\vert g_{\bq\lambda} \right\vert^2 \times 
    \bigg[ & N_\lambda^0 
    \delta(\varepsilon_{\bk'} - \varepsilon_\bk - \hbar\omega_\lambda) 
    \bigg. \nonumber \\
    & + \bigg. \left(1 + N_\lambda^0 \right) 
      \delta(\varepsilon_{\bk'} - \varepsilon_{\bk} +
        \hbar\omega_\lambda)
      \bigg]  .
\end{align}
For the in-scattering part, the desired $\cos{\theta_\bk}$ factor in
Eq.~\eqref{eq:I'} can be extracted from $\psi_{\bk'}$ using the relation
$\cos{\theta_{\bk'}} = \cos{\theta_\bk}\cos{\theta_{\bk\bk'}} -
\sin{\theta_\bk}\sin{\theta_{\bk\bk'}}$. Since the sine term vanishes from
symmetry consideration, the in-scattering part of the inelastic collision
integral reduces to
\begin{equation}
  \label{eq:I_in}
  {I'}_\text{inel}^\text{in}[\phi_k] 
  =  \sum_{\bk'} \phi_{k'} \cos \theta_{\bk\bk'} P_{\bk\bk'} 
                 \frac{1 - f_{\bk'}^0}{1 - f_\bk^0}   .
\end{equation}
The evaluation of the $\bk'$-sum in Eqs.~\eqref{eq:I_out} and~\eqref{eq:I_in} is
outlined in App.~\ref{app:collision}. Here, the assumption of dispersionless
optical phonons allows for an semianalytical treatment. For zero- and
first-order coupling within the deformation potential approximation, the sum can
be carried out analytically. The resulting expressions for the collision
integral are given in
Eqs.~\eqref{eq:I_out_zero},~\eqref{eq:I_in_zero},~\eqref{eq:I_out_first},~\eqref{eq:I_in_first1}
and~\eqref{eq:I_in_first2}. In the case of the Fr{\"o}hlich interaction, the
angular part of the $\bq$-integral must be done numerically.

\subsubsection{Optical deformation potential scattering rates}

In spite of the fact that the collision integral for inelastic scattering on
optical phonons cannot be recast in the form of a (momentum) relaxation
time~\cite{footnote3}, a scattering rate related to the inverse carrier
lifetime can still be defined from the out-scattering part of the inelastic
collision integral alone. The scattering rate so defined is given by
\begin{equation}
  \label{eq:tau_optical}
  \frac{1}{\tau_\bk^{\text{inel}}} = \sum_{\bk'} P_{\bk\bk'} 
       \frac{1 - f_{\bk'}^0}{1 - f_\bk^0} 
\end{equation}
and corresponds to the imaginary part of the electronic self-energy in the Born
approximation~\cite{SmithJensen}. Below, the resulting scattering rates for
zero-order and first-order deformation potential scattering are given for
non-degenerate carriers, i.e. with the Fermi factors in
Eq.~\eqref{eq:tau_optical} neglected. They follow straight-forwardly from the
expressions for the out-scattering part of the collision integral derived in
App.~\ref{app:collision}. It should be noted that the scattering rate for the
zero-order deformation potential interaction given below in
Eq.~\eqref{eq:tau_zero_optical}, in fact defines a proper momentum relaxation
time because the in-scattering part of the collision integral vanishes in this
case (see App.~\ref{app:collision}).

For zero-order deformation potential scattering, the scattering rate is
independent of the carrier energy and given by
\begin{align}
  \label{eq:tau_zero_optical}
  \frac{1}{\tau_\bk^\text{inel}} = \frac{m^* (D_\lambda^0)^2}{2\hbar^2\rho\omega_\lambda}
   \bigg[ 1 + e^{\hbar\omega_\lambda / k_\text{B}T} 
          \Theta(\varepsilon_{\bk} - \hbar\omega_{\lambda} )
   \bigg] N_{\lambda}^0 .
\end{align}
Here, $\Theta(x)$ is the Heavyside step function which assures that only
electrons with sufficient energy can emit a phonon. 

The scattering rate for coupling via the first-order deformation potential is
found to be
\begin{align}
  \label{eq:tau_first_optical}
  \frac{1}{\tau_\bk^\text{inel}} = \frac{m^{*2} (D_\lambda^1)^2}{\hbar^4\rho\omega_\lambda}
   \bigg[ &
       \left( 2\varepsilon_\bk + \hbar\omega_\lambda \right) 
       + e^{\hbar\omega_\lambda / k_\text{B}T}  
       \nonumber \\
     & \times
         \Theta(\varepsilon_{\bk} - \hbar\omega_{\lambda} )
         \left( 2\varepsilon_\bk - \hbar\omega_\lambda \right) 
   \bigg] N_\lambda^0 .
\end{align}
Due to the linear dependence on the carrier energy, zero-order scattering
processes dominate first-order processes at low energies. Only under high-field
conditions where the carriers are accelerated to high velocities will
first-order scattering become significant.

The expressions for the scattering rates in Eqs.~\eqref{eq:tau_zero_optical}
and~\eqref{eq:tau_first_optical} above apply to scattering on dispersionless
intervalley acoustic phonons and intra/intervalley optical phonons. Except for a
factor of $\sqrt{\varepsilon_\bk \pm \hbar\omega_\lambda}$ originating from the
density of states, the energy dependence of the scattering rates is identical to
that of their three-dimensional analogs~\cite{Ferry}.

\section{Results}
\label{sec:V}

In the following, the scattering rate and phonon-limited mobility in
single-layer MoS$_2$ are studied as a function of carrier energy, temperature
$T$ and carrier density $n$ using the material parameters collected in
Tab.~\ref{tab:parameters}. Here, the reported deformation potentials represent
effective coupling parameters for the deformation potential approximation in
Eqs.~\eqref{eq:M_acoustic} and~\eqref{eq:M_optical} (see below).
\begin{table}[!b]
\begin{ruledtabular}
\begin{tabular}{lcc}
Parameter &  Symbol  & Value  \\ 
\hline                                     
Lattice constant               &   $a$                     &   3.14 \AA            \\
Ion mass density               &   $\rho$                  &   $3.1\times 10^{-7}$ g/cm$^2$ \\
Effective electron mass        &   $m^*$                   &   0.48 $m_e$          \\
Transverse sound velocity      &   $c_\text{TA}$           &   $4.2 \times 10^3$ m/s  \\
Longitudinal sound velocity    &   $c_\text{LA}$           &   $6.7 \times 10^3$ m/s  \\
Acoustic deformation potentials&                                      &                  \\
TA                             &   $\Xi_\text{TA}$                    &   $1.6$ eV       \\
LA                             &   $\Xi_\text{LA}$                    &   $2.8$ eV       \\
Optical deformation potentials &                                      &                  \\
TA                             &  $D_{\mathbf{K},\text{TA}}^1$        &   $5.9$ eV       \\
LA                             &  $D_{\mathbf{K},\text{LA}}^1$        &   $3.9$ eV       \\
TO                             &  $D_{\mathbf{\Gamma},\text{TO}}^1$   &   $4.0$ eV       \\
TO                             &  $D_{\mathbf{K},\text{TO}}^1$        &   $1.9$ eV       \\
LO                             &  $D_{\mathbf{K},\text{LO}}^0$        &   $2.6 \times 10^8$ eV/cm  \\
Homopolar                      &  $D_{\mathbf{\Gamma},\text{HP}}^0$   &   $4.1 \times 10^8$ eV/cm  \\
Fr{\"o}hlich interaction (LO)  &                           &                       \\
Effective layer thickness      &   $\sigma$                &   4.41 \AA            \\
Coupling constant              &  $g_\text{Fr}$            &   $98$ meV            \\
Optical phonon energies        &                           &                       \\
Polar LO                       &  $\hbar\omega_\text{LO}$  &   48 meV              \\
Homopolar                      &  $\hbar\omega_\text{HP}$  &   50 meV              \\
\end{tabular}
\end{ruledtabular}
\caption{Material parameters for single-layer MoS$_2$. Unless otherwise stated,
  the parameters have been calculated from first-principles as described in the
  Secs.~II and~III. The $\mathbf{\Gamma}/\mathbf{K}$-subscript on the optical
  deformation potentials, indicate couplings to the intra/intervalley phonons.}
\label{tab:parameters}
\end{table}

\subsection{Scattering rates} 

\begin{figure}[!t]
  \includegraphics[width=0.75\linewidth]{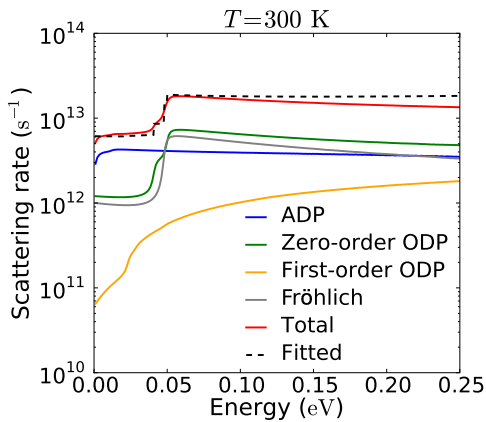}
  \caption{(Color online) Scattering rates for the different electron-phonon
    couplings as a function of carrier energy at $T=300$ K. The scattering rates
    have been calculated from the first-principles electron-phonon couplings in
    Figs.~\ref{fig:couplings} and~\ref{fig:frohlich}. The (black) dashed line
    shows the scattering rate obtained using the fitted deformation potential
    parameters defined in Eqs.~\eqref{eq:M_acoustic} and~\eqref{eq:M_optical}.
    The kinks in the curves for the optical scattering rates mark the onset of
    optical phonon emissions.}
\label{fig:scatteringrates}
\end{figure}

With access to the first-principles electron-phonon couplings, the scattering
rate taking into account the anisotropy and more complex $q$-dependence of the
first-principles coupling matrix elements can be evaluated. They are obtained
using the expression for the scattering rate in Eq.~\eqref{eq:tau_optical} with
the respective transition matrix elements for acoustic and optical phonon
scattering given in Eqs.~\eqref{eq:P_acoustic}
and~\eqref{eq:P_optical}~\cite{footnote4}. In order to account for the different
coupling matrix elements in the $K,K'$-valleys, the scattering rates have been
averaged over different high-symmetry directions of the carrier wave vector
$\bk$. The resulting scattering rates are shown in
Fig.~\ref{fig:scatteringrates} as a function of the carrier energy for
non-degenerate carriers at $T=300$ K. The different lines show the contributions
to the total scattering rate from the various electron-phonon couplings to the
intra and intervalley phonons which have been grouped according to their
coupling type, i.e. acoustic deformation potentials (ADPs), zero/first-order
optical deformation potentials (ODPs) and the Fr{\"o}hlich interaction. The
acoustic deformation potential scattering includes the quasielastic intravalley
scattering on the TA and LA phonons with linear dispersions. Scattering on
intervalley acoustic phonons is considered as optical deformation potential
scattering. Both the total scattering rate and the contributions from the
different coupling types have been obtained using Matthiessen's rule by summing
the scattering rates from the individual phonons, i.e.
\begin{equation}
  \label{eq:matthiessen}
  \tau_\text{tot}^{-1} = \sum_\lambda \tau_\lambda^{-1} .
\end{equation}

For carrier energies below the optical phonon frequencies, the total scattering
rate is dominated by acoustic deformation potential scattering. At higher
energies zero-order deformation potential scattering and polar optical
scattering via the Fr{\"o}hlich interaction become dominant. Due to the linear
dependence on the carrier energy, the first-order deformation potential
scattering on the intervalley acoustic phonons and the optical phonons is in
general only of minor importance for the low-field mobility. This is also the
case here, where it is an order of magnitude smaller than the other scattering
rates for almost the entire plotted energy range. The jumps in the curves for
the optical scattering rates at the optical phonon energies
$\hbar\omega_\lambda$ are associated with the threshold for optical phonon
emission where the carriers have sufficient energy to emit optical phonons

\subsection{Deformation potentials}

In this section, we determine the deformation potential parameters in
single-layer MoS$_2$. They can be used in the following study of the low-field
mobility within the Boltzmann approach outlined in Section~\ref{sec:IV}.

The energy dependence of the first-principles based scattering rates in
Fig.~\ref{fig:scatteringrates} to a high degree resembles that of the analytic
expressions for the deformation potential scattering rates in
Eqs.~\eqref{eq:tau_acoustic},~\eqref{eq:tau_zero_optical}
and~\eqref{eq:tau_first_optical}. For example, the acoustic and zero-order
deformation potential scattering rates are almost constant in the plotted energy
range. The first-principles electron-phonon couplings can therefore to a good
approximation be described by the simpler isotropic deformation potentials in
Eqs.~\eqref{eq:M_acoustic} and~\eqref{eq:M_optical}. The deformation potentials
are obtained by fitting the associated scattering rates for each of the intra
and intervalley phonons separately to the first-principles scattering rates.
The resulting deformation potential values are summarized in
Tab.~\ref{tab:parameters}. In analogy with deformation potentials extracted from
experimental mobilities, the theoretical deformation potentials represent
effective coupling parameters that implicitly account for the anisotropy and the
full $q$-dependence of the first-principles electron-phonon couplings. However,
as momentum and energy conservation limit phonon scattering to involve phonons
in the vicinity of the $\Gamma/K$-point~\cite{footnote2}, the fitting procedure
yields deformation potentials close to the direction averaged $\bq \rightarrow
\mathbf{\Gamma}$/$\mathbf{K}$ limiting behavior of the first-principles coupling
matrix elements. For the zero-order deformation potentials, the sampling of the
coupling matrix elements away from the $\Gamma$-point in the $\bk'$ sum in
Eq.~\eqref{eq:tau_optical}, leads to a deformation potentials which is slightly
smaller than the $\Gamma$-point value of the coupling matrix element.

The total scattering rate resulting from the fitted deformation potentials is
shown in Fig.~\ref{fig:scatteringrates} to give an almost excellent description
of the first-principles scattering rate. With the mobility in the relaxation
time approximation given by the energy-weighted average of the relaxation time
(see Eq.~\eqref{eq:mobility}), the difference between the associated mobilities
will be negligible. Hence, the deformation potentials provide well-founded
electron-phonon coupling parameters for low-field studies of the mobility.

\subsection{Mobility}

\begin{figure}[!t]
  \begin{minipage}{0.48\linewidth}
    \includegraphics[width=0.98\linewidth]{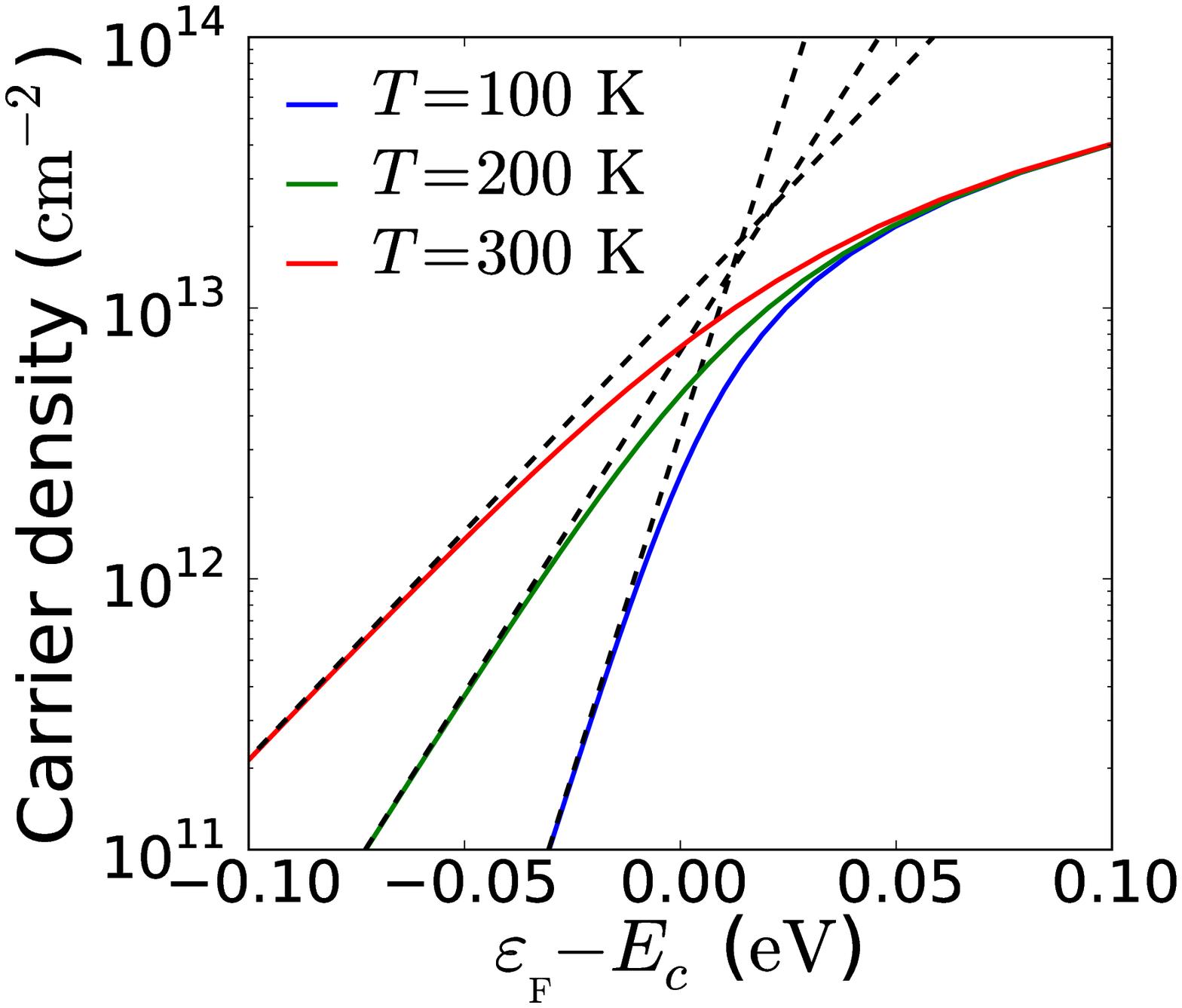}
  \end{minipage} \hfill
  \begin{minipage}{0.48\linewidth}
    \includegraphics[width=0.98\linewidth]{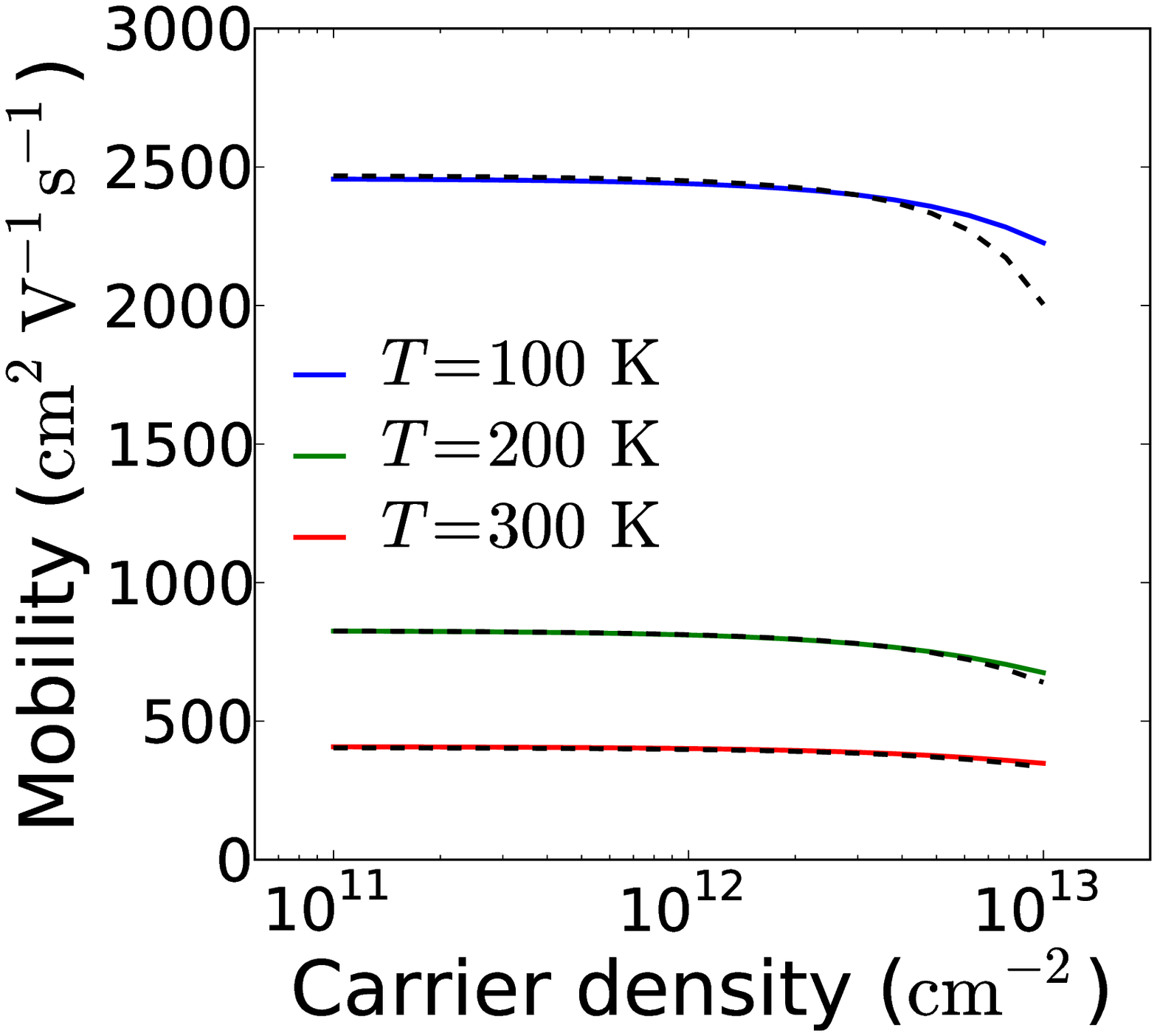}
  \end{minipage}
  \caption{(Color online) Left: Carrier density as a function of the Fermi
    energy for different temperatures. The Fermi energy is measured relative to
    the conduction band edge $E_c$ Right: Mobility as a function of carrier
    density for the same temperatures. In both plots, the (black) dashed lines
    show the results obtained with Boltzmann statistics and the relaxation time
    approximation (only right plot).}
\label{fig:mu_vs_n}
\end{figure}

The high density of states in the $K,K'$-valleys of the conduction band in
general results in non-degenerate carrier distributions in single-layer MoS$_2$.
This is illustrated in the left plot of Fig.~\ref{fig:mu_vs_n} which shows the
carrier density versus the position of the Fermi level for different
temperatures. At room temperature, carrier densities in excess of $\sim 8 \times
10^{12}$~cm$^{-2}$ are needed to introduce the Fermi level into the conduction
band and probe the Fermi-Dirac statistics of the carriers. Thus, only at the
highest reported carrier densities of $n \sim 10^{13}$~cm$^{-2}$~\cite{Geim:2D},
are the carriers degenerate at room temperature. For the lowest temperature
$T=100$~K, the transition to degenerate carriers occurs at a carrier density of
$\sim 2\times10^{12}$~cm$^{-2}$. The transition from non-degenerate to
degenerate distributions is also illustrated by the discrepancy between the full
and dashed lines which shows the Fermi level obtained with Boltzmann statistics.

The right plot of Fig.~\ref{fig:mu_vs_n} shows the phonon-limited drift mobility
calculated with the full collision integral as a function of carrier density for
the same set of temperatures. The dashed lines represent the results obtained
with Boltzmann statistics and the relaxation time approximation using the
expressions in Section~\ref{sec:IV} for optical phonon scattering and Matthiessen's
rule for the total relaxation time. The strong drop in the mobility from $\sim
2450$~cm$^2$~V$^{-1}$~s$^{-1}$ at $T = 100$~K to $\sim
400$~cm$^2$~V$^{-1}$~s$^{-1}$ at $T = 300$~K, is a consequence of the increased
phonon scattering at higher temperatures due to larger phonon population of, in
particular, optical phonons. The relatively low intrinsic room-temperature
mobility of single-layer MoS$_2$ can be attributed to both the significant
phonon scattering and the large effective mass of $0.48$ $m_e$ in the conduction
band.  While the mobility decreases strongly with increasing temperature, it is
relatively independent on the carrier density. The weak density dependence of
the mobility originates from the energy-weighted average in
Eq.~\eqref{eq:tau_averaged} where the derivative of the Fermi-Dirac distribution
changes slowly as a function of carrier density for non-degenerate carriers. As
the Fermi energy is introduced into the band with increasing carrier density,
the derivative of the Fermi-Dirac distribution to a larger extent probes the
scattering rate at higher energies leading to a decrease of the mobility. This
effect is most prominent at $T=100$~K where the level of degeneracy is larger
compared to higher temperatures.

Surprisingly, the relaxation time approximation is seen to work extremely well.
The deviation from the full treatment at high carrier densities stems from the
assumption of non-degenerate carriers and not the relaxation time approximation.
The reason for the good performance of the relaxation time approximation shall
be found in the in-scattering part of the collision integral in the full
treatment. As the in-scattering part of the collision integral vanishes for
zero-order deformation potential coupling and is small compared to the
out-scattering part otherwise, the full collision integral does not differ
significantly from the corresponding scattering rate as defined in
Eq.~\eqref{eq:tau_optical} of Section~\ref{sec:IV}, i.e. with the in-scattering
part neglected. The good performance of the relaxation time approximation
therefore seems to be of general validity for the phonon collision integral even
in the presence inelastic scattering on optical phonons.

Finally, we study the temperature dependence of the mobility in more detail. In
general, room temperature mobilities are to a large extent dominated by optical
phonon scattering. This is manifested in the temperature dependence of the
mobility which follows a $\mu \sim T^{-\gamma}$ law where the exponent $\gamma$
depends on the dominating scattering mechanism. For acoustic phonon scattering
above the Bloch-Gr{\"u}neisen temperature, the temperature dependence of the
scattering rate in Eq.~\eqref{eq:tau_acoustic} results in $\gamma=1$. At higher
temperatures where optical phonon scattering starts to dominate, the mobility
acquires a stronger temperature dependence with $\gamma > 1$. In this regime,
the exponent depends on the optical phonon frequencies and the electron-phonon
coupling strength. In an early study of the in-plane mobility of bulk
MoS$_2$~\cite{Mooser:Mobility}, the measured room-temperature exponent of
$\gamma\sim 2.6$ was found to be consistent with scattering on the homopolar
mode via a zero-order deformation potential.
\begin{figure}[!t]
  \includegraphics[width=0.75\linewidth]{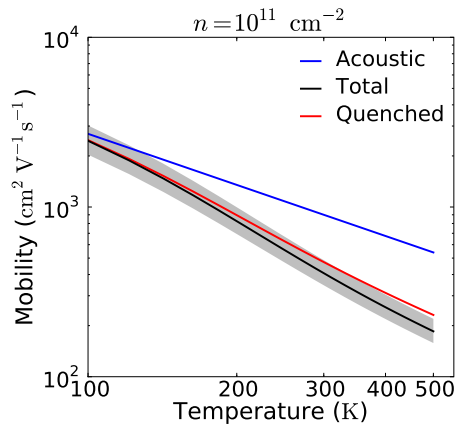}
  \caption{(Color online) Mobility vs temperature. For comparison, the mobility
    in the presence of only acoustic deformation potential scattering with the
    $\mu\sim T^{-1}$ temperature dependence is shown. The (gray) shaded area
    shows the variation in the mobility associated with a $10\%$ uncertainty in
    the calculated deformation potentials.}
\label{fig:mu_vs_T}
\end{figure}

In Fig.~\ref{fig:mu_vs_T}, we show the temperature dependence of the mobility at
$n=10^{11}$~cm$^{-2}$ calculated with the full collision integral. For
comparison, the mobility limited by acoustic phonon scattering with the $\mu
\sim T^{-1}$ temperature dependence is also included. To illustrate the effect
of an uncertainty in the calculated deformation potentials, the shaded area
shows the variation of the mobility with a change in the deformation potentials
of $\pm 10 \%$. The calculated room-temperature mobility of $\sim
410$~cm$^2$~V$^{-1}$~s$^{-1}$ is in fair agreement with the recently reported
experimental value of $\sim 200$~cm$^2$~V$^{-1}$~s$^{-1}$ in the top-gated
sample of Ref.~\onlinecite{Kis:MoS2Transistor} where additional scattering
mechanisms as e.g. impurity and surface-optical phonon scattering must be
expected to exist. Over the plotted temperature range, the mobility undergoes a
transition from being dominated by acoustic phonon scattering at $T=100$~K to
being dominated by optical phonon scattering at higher temperatures with a
characteristic exponent of $\gamma > 1$. At room temperature, the mobility
follows a temperature dependence with $\gamma=1.69$. The increase in the
exponent is almost exclusively due to the optical zero-order deformation
potential couplings and the Fr{\"o}hlich interaction while the first-order
deformation potential couplings contribute only marginally. The exponent found
here is considerably lower than the above-mentioned exponent of $\gamma\sim 2.6$
for bulk MoS$_2$, indicating that the electron-phonon coupling in bulk and
single-layer MoS$_2$ differ. Indeed, the transition from an indirect band gap in
bulk MoS$_2$ to a direct band gap in single-layer MoS$_2$ which shifts the
bottom of the conduction band from the valley located along the $\Gamma$-K path
to the $K,K'$-valleys, could result in a change in the electron-phonon coupling.

In top-gated samples as the one studied in Ref.~\onlinecite{Kis:MoS2Transistor},
the sandwiched structure with the MoS$_2$ layer located between substrate and
gate dielectric, is likely to result in a quenching of the characteristic
homopolar mode which is polarized in the direction normal to the layer. To
address the consequence of such a quenching, the mobility in the absence of the
zero-order deformation potential originating from the coupling to homopolar mode
is also shown in Fig.~\ref{fig:mu_vs_T}. Here, the curve for the quenched case
shows a decrease in the characteristic exponent to $\gamma=1.52$ and an increase
in the mobility of $\sim 70$~cm$^2$~V$^{-1}$~s$^{-1}$ at room temperature.
Despite the significant deformation potential of the homopolar mode, the effect
of the quenching on the mobility is minor.

\section{Discussions and conclusions}
\label{sec:VI}

Based on our finding for the phonon-limited mobility in single-layer MoS$_2$, it
seems likely that the low experimental mobilities of $\sim
1$~cm$^2$~V$^{-1}$~s$^{-1}$ reported in
Refs.~\onlinecite{Geim:2D,Fuhrer:Ultrathin,Kis:MoS2Transistor} are dominated by
other scattering mechanisms as e.g. charged impurity scattering. The main reason
for the increase in the mobility to 200~cm$^2$~V$^{-1}$~s$^{-1}$ observed when
depositing a high-$\kappa$ dielectric on top of the MoS$_2$ layer in
Ref.~\onlinecite{Kis:MoS2Transistor} is therefore likely to be impurity
screening. The quenching of the out-of-plane homopolar phonon only led to a
minor increase in the mobility and cannot alone explain the above-mentioned
increase in the mobility. It may, however, also contribute. A comparison of the
temperature dependence of the mobility in samples with and without the top-gate
structure can help to clarify the extent to which a quenching of optical phonons
contributes to the experimentally observed mobility increase. With our
theoretically predicted room-temperature mobility of
410~cm$^2$~V$^{-1}$~s$^{-1}$, the observed enhancement in the mobility to
200~cm$^2$~V$^{-1}$~s$^{-1}$ induced by the high-$\kappa$ dielectric, suggests
that dielectric engineering~\cite{Konar:Engineering} is an effective route
towards phonon-limited mobilities in 2D materials via efficient screening of
charge impurities.

A rough estimate of the impurity concentrations required to dominate phonon
scattering can be inferred from the phonon scattering rate of $\tau^{-1} \sim
10^{13}$~s$^{-1}$ in Fig.~\ref{fig:scatteringrates}. The scattering rate is
related to the mean-free path $\lambda$ of the carriers via $\lambda = \tau
\expect{v}$ where $\expect{v}$ is their mean velocity. Using the velocity of the
mean energy carriers for a non-degenerate distribution where
$\expect{\varepsilon_\bk} = k_\text{B} T$, we find a mean-free path of $\lambda
\sim 14$~nm at $T=300$~K. In order for impurity scattering to dominate, the
impurity spacing must be on the order of the phonon mean-free path or smaller.
This results in a minimum impurity concentration of $\sim
5\times10^{11}$~cm$^{-2}$ in order to dominate phonon scattering. The high value
of the estimated impurity concentration needed to dominate phonon scattering is
in agreement with the experimental observation that low-mobility single-layer
MoS$_2$ samples are heavily doped semiconductors~\cite{Geim:2D}.

As discussed in Section~\ref{sec:II}, independent GW quasi-particle
calculations~\cite{Ciraci:Functionalization,Olsen:MoS2} suggest that the
ordering of the valleys at the bottom of the conduction band might not be as
clear as predicted by DFT. If this is the case, the satellite valleys inside the
Brillouin zone must also be taken into account in the solution of Boltzmann
equation. This gives rise to additional intervalley scattering channels that
together with the larger average effective electron mass $m^* \sim 0.6$ $m_e$ of
the satellite valley will result in mobilities below the values predicted here.

In conclusion, we have used a first-principles based approach to establish the
strength and nature of the electron-phonon coupling and calculate the intrinsic
phonon-limited mobility in single-layer MoS$_2$. The calculated room-temperature
mobility of 410~cm$^2$~V$^{-1}$~s$^{-1}$ is to a large extent dominated by
optical deformation potential scattering on the intravalley homopolar and
intervalley LO phonons as well as polar optical scattering on the intravalley LO
phonon via the Fr{\"o}hlich interaction. The mobility follows a $\mu \sim
T^{-1.69}$ temperature dependence at room temperature characteristic of optical
phonon scattering. A quenching of the homopolar mode likely to occur in
top-gated samples, results in a change of the exponent in the temperature
dependence of the mobility to 1.52. With effective masses, phonons and measured
mobilities in other semiconducting metal dichalcogenides being similar to those
of MoS$_2$~\cite{Mooser:Mobility,Fivaz:Dimensionality}, room-temperature
mobilities of the same order of magnitude can be expected in their single-layer
forms.

\begin{acknowledgments}
  The authors would like to thank O. Hansen and J. J. Mortensen for illuminating
  discussions. KK has been partially supported by the Center on Nanostructuring
  for Efficient Energy Conversion (CNEEC) at Stanford University, an Energy
  Frontier Research Center funded by the U.S. Department of Energy, Office of
  Science, Office of Basic Energy Sciences under Award Number DE-SC0001060. CAMD
  is supported by the Lundbeck Foundation.
\end{acknowledgments}

\appendix

\section{First-principles calculation of the electron-phonon interaction}
\label{app:e-ph}

The first-principles scheme for the calculation of the electron-phonon
interaction used in this work is outlined in the following. Contrary to other
first-principles approaches for the calculation of the electron-phonon
interaction which are based on pseudo
potentials~\cite{Sjakste:GaAsInter,Louie:e-ph,Sjakste:Silicon,Kim:ElPhGraphene},
the present approach is based on the PAW method~\cite{Blochl:PAW} and is
implemented in the GPAW DFT package~\cite{GPAW,GPAW1,GPAW2}.

Within the adiabatic approximation for the electron-phonon interaction, the
electron-phonon coupling matrix elements for the phonon $\bq\lambda$ and Bloch
state $\bk$ can be expressed as
\begin{equation}
  \label{eq:g_bloch}
  g^{\lambda}_{\bk\bq}  = 
      \sqrt{\frac{\hbar}{2MN\omega_{\bq\lambda}}}
      \sum_{\alpha}
      \bra{\mathbf{k+q}} \hat{\mathbf{e}}_{\alpha}^{\bq\lambda} \cdot
      \nabla_{\alpha}^{\bq} V(\br) 
      \ket{\bk} ,
\end{equation}
where $N$ is the number of unit cells, $M$ is an appropriately defined effective
mass, $\omega_{\bq\lambda}$ is the phonon frequency, and the $\bq$-dependent
derivative of the effective potential for a given atom $\alpha$ is defined by
\begin{equation}
  \label{eq:V_q}
  \nabla_\alpha^{\bq} V(\br) 
    = \sum_l e^{i \bq \cdot \mathbf{R}_l } 
      \nabla_{\alpha l} V(\br) ,
\end{equation}
where the sum is over unit cells in the system and $\nabla_{\alpha l} V$ is the
gradient of the potential with respect to the position of atom $\alpha$ in cell
$\mathbf{R}_l$. The polarization vectors $\hat{\mathbf{e}}_{\bq\lambda}$ of the
phonons appearing in Eq.~\eqref{eq:g_bloch} are normalized according to
$\sum_{\alpha} (M_{\alpha}/M)\vert \hat{\mathbf{e}}_{\alpha}^{\bq\lambda}
\vert^2 = 1$.

In order to evaluate the matrix element in Eq.~\eqref{eq:g_bloch}, the Bloch
states are expanded in LCAO basis orbitals $\ket{i \mathbf{R}_n}$ where
$i=(\alpha,\mu)$ is a composite index for atomic site and orbital index and
$\mathbf{R}_n$ is the lattice vector to the $n$'th unit cell. In the LCAO basis,
the Bloch states are expanded as $\ket{\bk} = \sum_i c_i^\bk \ket{i \bk}$ where
\begin{equation}
  \label{eq:bloch_sum}
  \ket{i \bk} = \frac{1}{N} \sum_n 
                e^{i \bk \cdot \mathbf{R}_n}
                \ket{i \mathbf{R}_n} .
\end{equation}
are Bloch sums of the localized orbitals. Inserting the Bloch sum expansion of
the Bloch states in the matrix element in Eq.~\eqref{eq:g_bloch}, the matrix
element can be written
\begin{align}
  \label{eq:g_blochsum}
  \bra{\mathbf{k+q}} \hat{\mathbf{e}}_{\bq\lambda} \cdot 
  & \nabla_\bq V(\br) \ket{\bk} 
     =  \sum_{ij} c_i^{*} c_j  \bra{i \mathbf{k+q}} \hat{\mathbf{e}}_{\bq\lambda} \cdot 
         \nabla_\bq V(\br) \ket{j \bk} \nonumber \\
     = & \frac{1}{N^2} \sum_{ij} c_i^{*} c_j  \sum_{lmn}
         e^{i (\bk + \mathbf{G}) \cdot (\mathbf{R}_n - \mathbf{R}_m) - 
            i \bq \cdot (\mathbf{R}_m - \mathbf{R}_l ) } \nonumber \\
     &  \quad\quad \times \bra{i \mathbf{R}_m} \hat{\mathbf{e}}_{\bq\lambda} \cdot 
         \nabla_l V(\br) \ket{j \mathbf{R}_n}
\end{align}
where the $\bk$-labels on the expansion coefficients have been discarded for
brevity. The phase factor from the reciprocal lattice
vector $\mathbf{G}$ can be neglected since $\mathbf{R}_n \cdot \mathbf{G} =
2\pi N$. By exploiting the translational invariance of the crystal the matrix
elements can be obtained from the gradient in the primitive cell as
\begin{align}
  & \bra{i \mathbf{R}_m} \hat{\mathbf{e}}_{\bq\lambda} \cdot \nabla_l V(\br) \ket{j \mathbf{R}_n}
  \nonumber \\
  & = \bra{i \mathbf{R}_m - \mathbf{R}_l} \hat{\mathbf{e}}_{\bq\lambda} \cdot 
      \nabla_0 V(\br) \ket{j \mathbf{R}_n - \mathbf{R}_l}
\end{align}
by performing the change of variables $\br' = \br - \mathbf{R}_l$ in the
integration. Inserting in Eq.~\eqref{eq:g_blochsum} and changing the summing
variables to $\mathbf{R}_m - \mathbf{R}_l$ and $\mathbf{R}_n - \mathbf{R}_l$ we
find for the matrix element
\begin{widetext}
\begin{align}
  \label{eq:g_blochsum_final}
  \bra{\mathbf{k+q}} \hat{\mathbf{e}}_{\bq\lambda} \cdot
  \nabla_\bq V(\br) \ket{\bk}
  & = \frac{1}{N^2} \sum_{ij} c_i^{*} c_j  \sum_{lmn}
      e^{i \bk \cdot (\mathbf{R}_n - \mathbf{R}_m) -
         i \bq \cdot \mathbf{R}_m}
       \bra{i \mathbf{R}_m} \hat{\mathbf{e}}_{\bq\lambda} \cdot
       \nabla_0 V(\br) \ket{j \mathbf{R}_n} \nonumber \\
  & = \frac{1}{N} \sum_{ij} c_i^{*} c_j  \sum_{mn}
     e^{i \bk \cdot (\mathbf{R}_n - \mathbf{R}_m) -
        i \bq \cdot \mathbf{R}_m}
      \bra{i \mathbf{R}_m} \hat{\mathbf{e}}_{\bq\lambda} \cdot
      \nabla_0 V(\br) \ket{j \mathbf{R}_n}
\end{align}
\end{widetext}
Here, the sum over $k$ in the first equality produces a factor of $N$.
The result for the matrix element in Eq.~\eqref{eq:g_blochsum_final} is similar
to the matrix element reported in Eq.~(22) of Ref.~\onlinecite{Louie:e-ph}
where a Wannier basis was used instead of the LCAO basis.

From the last equality in Eq.~\eqref{eq:g_blochsum_final}, the procedure for a
supercell-based evaluation of the matrix element has emerged. The matrix element
in the last equality involves the gradients of the effective potential
$\nabla_0 V$ with respect to atomic displacement in the reference cell at
$\mathbf{R}_0$. These can be obtained using a finite-difference approximation
for the gradient where the individual components are obtained as
\begin{equation}
  \label{eq:fd}
  \frac{\partial V}{\partial R_{\alpha i}}  = 
    \frac{V^+_{\alpha i}(\br) - V^-_{\alpha i}(\br)}{2\delta} .
\end{equation}
Here, $V^{\pm}_{\alpha i}$ denotes the effective potential with atom $\alpha$
located at $\mathbf{R}_\alpha$ in the reference cell displaced by $\pm \delta$
in direction $i=x,y,z$. The calculation of the gradient thus amounts to carrying
out self-consistent calculations for six displacements of each atom in the
primitive unit cell. Having obtained the gradients of the effective potential,
the matrix elements in the LCAO basis of the supercell must be calculated and
the sums over unit cells and atomic orbitals in Eq.~\eqref{eq:g_blochsum_final}
can be evaluated.

In general, the matrix elements of the electron-phonon interaction must be converged with
respect to the supercell size and the LCAO basis. In particular, since the
supercell approach relies on the gradient of the effective potential going to
zero at the supercell boundaries, the supercell must be chosen large enough that
the potential at the boundaries is negligible. For polar materials where the
coupling to the polar LO phonon is long-ranged in nature, a correct description
of the coupling can only be obtained for phonon wave vectors corresponding to
wavelengths smaller than the supercell size. Large supercells are therefore
required to capture the long-wavelength limit of the coupling constant (see e.g.
main text).

\subsection{PAW details}

In the PAW method, the effective single-particle DFT Hamiltonian is given by
\begin{equation}
  \label{eq:H_eff_PAW}
  H = -\frac{1}{2} \nabla^2 + v_\text{eff}(\br)
    + \sum_\alpha \sum_{i_1 i_2} \; \ket{\tilde{p}_{i_1}^\alpha} 
                                \Delta H_{i_1 i_2}^\alpha
                                \bra{\tilde{p}_{i_2}^\alpha} ,
\end{equation}
where the first term is the kinetic energy, $v_\text{eff}$ is the effective
potential containing contributions from the atomic potentials and the Hartree
and exchange-correlation potentials and the last term is the non-local part
including the atom-centered projector functions $\tilde{p}_i^\alpha(\br)$ and
the atomic coefficients $\Delta H_{i_1 i_2}^\alpha$. In contrast to
pseudo-potential methods where the atomic coefficients are constants, they
depend on the density and thereby also on the atomic positions in the PAW
method.

The diagonal matrix elements (i.e. $\bq = \mathbf{0}$) in
Eq.~\eqref{eq:g_blochsum} correspond to the first-order variation in the band
energy $\varepsilon_\bk$ upon an atomic displacement in the normal mode
direction $\hat{\mathbf{e}}_{\bq\lambda}$. Together with the matrix elements for
$\bq \neq \mathbf{0}$, these can be obtained from the gradient of the PAW
Hamiltonian with respect to atomic displacements. The derivative
of~\eqref{eq:H_eff_PAW} with respect to a displacement of atom $\alpha$ in
direction $i$ results in the following four terms
\begin{align}
  \frac{\partial H}{\partial R_{\alpha i}} = 
  \frac{\partial v_\text{eff}}{\partial R_{\alpha i}} +
  \sum_{i_1,i_2} 
  \bigg[ & 
    \ket{\frac{\partial\tilde{p}_{i_1}^\alpha}{\partial R_{\alpha i}}}
    \Delta H_{i_1 i_2}^\alpha
    \bra{\tilde{p}_{i_2}^\alpha} 
    \nonumber \\
  & + \ket{\tilde{p}_{i_1}^\alpha} 
    \Delta H_{i_1 i_2}^\alpha
    \bra{\frac{\partial\tilde{p}_{i_2}^\alpha}{\partial R_{\alpha i}}}
    \nonumber \\
  & + \ket{\tilde{p}_{i_1}^\alpha}
    \frac{\partial \Delta H_{i_1 i_2}^\alpha}{\partial R_{\alpha i}}
    \bra{\tilde{p}_{i_2}^\alpha}
  \bigg] .
\end{align}
While the gradient of the projector functions in the first and second terms
inside the square brackets can be evaluated analytically, the gradients of the
effective potential and the projector coefficients in the last term are obtained
using the finite-difference approximation in Eq.~\eqref{eq:fd}. 

Once the gradient of the PAW Hamiltonian has been obtained, the matrix elements
in Eq.~\eqref{eq:g_blochsum} can be obtained. Under the assumption that the
smooth pseudo Bloch wave functions $\tilde{\psi}_\bk$ from the PAW formalism is
a good approximation to the true quasi particle wave function, the matrix
element follows as
\begin{equation}
  \bra{\mathbf{k+q}} \hat{\mathbf{e}}_{\bq\lambda} \cdot 
  \nabla_\bq V(\br) \ket{\bk}  =
  \bra{\tilde{\psi}_\mathbf{k+q}} \hat{\mathbf{e}}_{\bq\lambda} 
  \cdot \nabla_\bq H \ket{\tilde{\psi}_\bk} ,
\end{equation}
where $\nabla_\bq H$ is given in Eq.~\eqref{eq:V_q}. With the pseudo wave
function expanded in an LCAO basis, the matrix element is evaluated following
Eq.~\eqref{eq:g_blochsum_final}.

In order to verify the calculated matrix elements, we have carried out
self-consistent calculations of the variation in the band energies with respect
to atomic displacements along the phonon normal mode. For the coupling to the
$\Gamma$-point homopolar phonon in single-layer MoS$_2$, we find that the
calculated matrix element of 4.8~eV/{\AA} agrees with the self-consistently
calculated value to within $0.1$~eV/{\AA}.

As the matrix elements of the electron-phonon interaction are sensitive to the
behavior of the potential and wave functions in the vicinity of the atomic
cores, variations in electron-phonon couplings obtained with different
pseudo-potential approximations can be expected. We have confirmed this via
self-consistent calculations of the band energy variations, which showed that
the coupling to the homopolar $\Gamma$-point phonon increases with
$0.3$~eV/{\AA} compared to the PAW value when using norm-conserving HGH pseudo
potentials.

\section{Fr{\"o}hlich interaction in 2D materials}
\label{app:frohlich}

In the long-wavelength limit, the lattice vibration of the polar LO mode gives
rise to a macroscopic polarization that couples to the charge carriers. In
three-dimensional bulk systems, the coupling strength is given by
Eq.~\ref{eq:frohlich} which diverges as $1/q$. Using dielectric continuum
models, the coupling has been studied for confined carriers and LO phonons in
semiconductor heterostructures~\cite{Ando:ElphHetero,Lugli:Microscopic} where a
non-diverging interaction is found. For atomically thin materials, however,
macroscopic dielectric models are inappropriate. Using an approach based on the
atomic Born effective charges $Z^*$, a functional form that fits the calculated
values of the Fr{\"o}hlich interaction in Fig.~\ref{fig:frohlich} is here
derived.

\subsection{Polarization field and potential}

In two-dimensional materials, the polarization from the lattice vibration of the
polar LO phonon is oriented along the plane of the layer. It can be expressed in
terms of the relative displacement $\mathbf{u}_\bq$ of the unit cell atoms as
\begin{equation}
  \mathbf{P}_\bq(z) = \frac{Z^*}{\varepsilon_\infty A} \mathbf{u}_\bq f_\bq(z) ,
\end{equation}
where $\bq$ is the two-dimensional phonon wave vector, $\varepsilon_\infty$ is
the optical dielectric constant, $Z^*$ is the Born effective charge of the atoms
(here assumed to be the same for all atoms) and $f_\bq$ describes the profile of
the polarization in the direction perpendicular (here the $z$-direction) to the
layer. The associated polarization charge $\rho = - \nabla \cdot \mathbf{P}$ is
given by $\rho_\bq = -i \bq \cdot \mathbf{P}_\bq$. The resulting scalar
potential which couples to the carriers follows from Poisson's equation. Fourier
transforming in all three directions, Poisson's equation takes the form
\begin{equation}
  \label{eq:poisson_ft}
  (q^2 + k^2) \phi_\bq (k) 
  = -i \frac{Z^*}{\varepsilon_\infty A} \bq \cdot \mathbf{u}_\bq f_\bq(k) ,
\end{equation}
where $k$ is the Fourier variable in the direction perpendicular to the plane of
the layer and $\bq \parallel \mathbf{u}_\bq$ for the LO phonon.

In 2D materials, the $z$-profile of the polarization field can to a good
approximation be described by a $\delta$-function, i.e.
\begin{equation}
  f_\bq(z) = \delta(z) .
\end{equation}
Inserting the Fourier transform $f_\bq(k) = 1$ in Eq.~\eqref{eq:poisson_ft},
we find for the potential
\begin{align}
  \phi_\bq(z) & = -i \frac{Z^*u_\bq}{\varepsilon_\infty A}
                \int \! dk \; e^{ikz} f_\bq(k)
                \frac{q}{q^2 + k^2} \nonumber \\
              & = -i \frac{Z^*u_\bq}{\varepsilon_\infty A}
                    e^{-q \abs{z}} 
\end{align}
which in agreement with the findings of
Refs.~\onlinecite{Ando:ElphHetero,Lugli:Microscopic} does not diverge in the
long-wavelength limit.

\subsection{Electron-phonon interaction}

In three-dimensional bulk systems, the $q$-dependence of the Fr{\"o}hlich
interaction is given entirely by the $1/q$-divergence of the potential
associated with the lattice vibration of the polar LO phonon. However, in two
dimensions, the interaction follows by integrating the potential
with the square of the envelope function $\chi(z)$ of the electronic Bloch
state. Hence, the Fr{\"o}hlich interaction is given by the matrix element 
\begin{equation}
  g_\text{LO} (q) = \int \! dz \; \chi^*(z) \phi_\bq(z) \chi (z) .
\end{equation}
For simplicity, we here assume a Gaussian profile for the envelope function
\begin{equation}
  \chi(z) = \frac{1}{\pi^{1/4} \sqrt{\sigma}} 
             e^{-z^2/2\sigma^2} ,
\end{equation}
where $\sigma$ denotes the effective width of the electronic Bloch function.
With this approximation for the envelope function, the Fr{\"o}hlich interaction
becomes
\begin{align}
   g_\text{LO}(q) & = g_\text{Fr} \times 
                      e^{q^2 \sigma^2 / 4} \text{erfc}(q \sigma / 2)
                      \nonumber \\
                  & \approx g_\text{Fr} \times \text{erfc}(q \sigma / 2) 
\end{align}
where $g_\text{Fr}$ is the Fr{\"o}hlich coupling constant, $\text{erfc}$ is the
complementary error function and the last equality holds in the long-wavelength
limit where $q^{-1} \gg \sigma$. As shown in Fig.~\ref{fig:frohlich} in the main
text, this functional form for the Fr{\"o}hlich interaction gives a perfect fit
to the calculated electron-phonon coupling for the polar LO mode.

\section{Evaluation of the inelastic collision integral}
\label{app:collision}

Following Eq.~\eqref{eq:I_inel_out_in} in the main text, the inelastic collision
integral for scattering on optical phonons is split up in separate out- and
in-scattering contributions,
\begin{equation}
  {I'}_\text{inel}[\phi_k] 
  = {I'}_\text{inel}^\text{out}[\phi_k] + 
    {I'}_\text{inel}^\text{in}[\phi_k]  ,
\end{equation}
which are given by Eq.~\eqref{eq:I_out} and~\eqref{eq:I_in}, respectively.
With the assumption of dispersionless optical phonons, the Fermi-Dirac and
Bose-Einstein distribution functions do not depend on the phonon wave vector and
can thus be taken outside the $\bq$-sum.

The out-scattering part of the collision integral then takes the form
\begin{align}
  \label{eq:I_out_1}
  \quad & {I'}_\text{inel}^\text{out}[\phi_k]  =  - \frac{2\pi}{\hbar} 
  \frac{\phi_k}{{1 - f_\bk^0}} \nonumber \\
  & \times
  \Bigg[
    N_\lambda^0  (1 - f_{\bk + \bq}^0) 
    \sum_\bq \left\vert g_\bq \right\vert^2 
    \delta(\varepsilon_{\bk + \bq} - \varepsilon_\bk - \hbar\omega_\lambda) 
    \nonumber \\
  & + (1 + N_\lambda^0 ) (1 - f_{\bk - \bq}^0) 
    \sum_\bq \left\vert g_\bq \right\vert^2 
    \delta(\varepsilon_{\bk - \bq} - \varepsilon_{\bk} + \hbar\omega_\lambda)
    \Bigg] .
\end{align}
For the in-scattering part, the additional $\cos\theta_{\bk,\bk\pm\bq}$ factor
is rewritten as
\begin{equation}
  \label{eq:cos_k_kplusq}
  \cos\theta_{\bk,\bk\pm\bq} = \frac{k \pm q \cos{\theta_{\bk,\bq}}}
                                    {k\sqrt{1 \pm \frac{\hbar\omega_\lambda}
                                                       {\varepsilon_\bk}}} ,
\end{equation}
leading to the following general form of the in-scattering part of the collision
integral
\begin{align}
  \label{eq:I_in_1}
  {I'}_\text{inel}^\text{in}& [\phi_k]  =  \frac{2\pi}{\hbar} 
  \frac{1}{{1 - f_\bk^0}} 
  \nonumber \\
  & \times
  \Bigg[
    N_\lambda^0  (1 - f_{\bk + \bq}^0) 
    \frac{\phi_{k+q}}{k \sqrt{1 + \frac{\hbar\omega_\lambda}{\varepsilon_\bk}}}
  \nonumber \\
  & \quad \times
  \sum_\bq \left\vert g_\bq \right\vert^2 (k + q \cos{\theta_{\bk,\bq}} ) 
  \delta(\varepsilon_{\bk + \bq} - \varepsilon_\bk - \hbar\omega_\lambda) 
    \nonumber \\
  & \quad 
    + (1 + N_\lambda^0 ) (1 - f_{\bk - \bq}^0)
      \frac{\phi_{k-q}}{k \sqrt{1 - \frac{\hbar\omega_\lambda}{\varepsilon_\bk}}}
    \nonumber \\
  & \quad \times
    \sum_\bq \left\vert g_\bq \right\vert^2 
    (k - q \cos{\theta_{\bk,\bq}} ) 
    \delta(\varepsilon_{\bk - \bq} - \varepsilon_{\bk} + \hbar\omega_\lambda)
    \Bigg]  .
\end{align}
This depends on the deviation function in $\phi_{k\pm q} \equiv
\phi(\varepsilon_\bk \pm \hbar\omega_\lambda)$ and thus couples the deviation
function at the initial energy $\varepsilon_\bk$ of the carrier to that at
$\varepsilon_\bk \pm \hbar\omega_\lambda$. The two terms inside the square
brackets account for emission and absorption out of the state $\bk \pm \bq$ and
into the state $\bk$, respectively.

\subsection{Integration over $q$}

The in- and out-scattering contributions to the inelastic part of the collision
integral in Eqs.~\eqref{eq:I_out_1} and~\eqref{eq:I_in_1} can be written on the
general form
\begin{equation}
  \label{eq:collision_general}
  {I'}_\text{inel}^\text{in/out}(\bk) = \sum_\bq f(q) 
  \delta(\varepsilon_{\bk \pm \bq} - \varepsilon_{\bk} \mp
  \hbar\omega_{\bq\lambda}) ,
\end{equation}
where the function $f$ accounts for the $q$-dependent functions inside the
$\bq$-sum. The
conservation of energy and momentum in a scattering event is secured by the
$\delta$-function entering the collision integral.

Following the procedure outlined in Ref.~\onlinecite{Mooser:Mobility}, the
integral over $q$ resulting from the $\bq$-sum in the collision integral is
evaluated using the following property of the $\delta$-function,
\begin{equation}
  \int \! dq \; F(q) \delta\left[ g(q) \right] = \sum_n
  \frac{F(q_n)}{\abs{g'(q_n)}} .
\end{equation}
Here, $F(q)=qf(q)$ and $q_n$ are the roots of $g$. Rewriting the argument of the
$\delta$-function in Eq.~\eqref{eq:collision_general} as
\begin{align}
  \label{eq:conservation}
  \varepsilon_{\bk \pm \bq} - \varepsilon_{\bk} \mp \hbar\omega_{\bq\lambda}
  & = \frac{\hbar^2 (\bk \pm \bq)^2}{2m^*}
      - \frac{\hbar^2k^2}{2m^*} \mp \hbar\omega_{\bq\lambda}
  \nonumber \\
  & = \frac{\hbar^2}{2 m^*} 
    \left(q^2 \pm 2 kq \cos{\theta_{\bk,\bq}} \mp \frac{2 m^*\omega_{\bq\lambda}}{\hbar} 
    \right)
  \nonumber \\
  & \equiv \frac{\hbar^2}{2 m^*} g(q)
\end{align}
we get
\begin{equation}
  \label{eq:collision_delta}
  {I'}(\bk) = \frac{A}{(2\pi)^2}
  \int \! d\theta_{\bk,\bq} \; \frac{2 m^*}{\hbar^2} \sum_n \frac{q_n f(q_n)}{\abs{g'(q_n)}} .
\end{equation}
Depending on the $q$-dependence of the phonon frequency, different solutions for
the roots $q_n$ result.

\subsubsection{Dispersionless optical phonons}

With the assumption of dispersionless optical phonons, i.e. $\omega_{\bq\lambda}
= \omega_\lambda$, the roots of $g$ in Eq.~\eqref{eq:conservation} become
\begin{equation}
  \label{eq:roots}
  q / k = \mp \cos{\theta_{\bk,\bq}} \pm 
          \sqrt{\cos^2{\theta_{\bk,\bq}} \pm \frac{\hbar\omega_\lambda}{\varepsilon_\bk}}
\end{equation}
for absorption (upper sign in the first and last term) and emission (lower sign
in the first and last term), respectively. 

For absorption there is only one positive root $q_+^a$ given by the plus sign in
front of the square root,
\begin{equation}
  \label{eq:q^abs}
  q_+^a = -k \cos{\theta_{\bk,\bq}} +
           k \sqrt{\cos^2{\theta_{\bk,\bq}} + \frac{\hbar\omega_\lambda}{\varepsilon_\bk}}
\end{equation}
where all values of $\theta_{\bk,\bq}$ are allowed. The absorption terms in the
collision integral then becomes
\begin{equation}
  {I'}_a(\bk) = \frac{A}{(2\pi)^2}
  \int_0^{2\pi} \! d\theta_{\bk,\bq} \; \frac{m^*}{k\hbar^2} 
  \frac{q_+^a f(q_+^a)}
  {\sqrt{\cos^2{\theta_{\bk,\bq}} + \frac{\hbar\omega_\lambda}{\varepsilon_\bk}}}.
\end{equation}
In the case of phonon emission, both possible roots in Eq.~\eqref{eq:roots}
\begin{equation}
  \label{eq:q^emis}
  q_\pm^e =  k \cos{\theta_{\bk,\bq}} \pm
         k \sqrt{\cos^2{\theta_{\bk,\bq}} - \frac{\hbar\omega_\lambda}{\varepsilon_\bk}}
\end{equation}
can take on positive values. However, the minus sign inside the square root
restricts the allowed values of the integration angle to the range
$\theta_{\bk,\bq} \le \theta_0 =
\arccos{(\hbar\omega_\lambda/\varepsilon_\bk)^{1/2}}$. Furthermore, in order to
secure a positive argument to the square root, the electron energy must be
larger than the phonon energy, $\varepsilon_\bk > \hbar\omega_\lambda$. Hence,
the emission terms can be obtained as
\begin{align}
  {I'}_e(\bk) = & \frac{A}{(2\pi)^2} \Theta(\varepsilon_\bk - \hbar\omega_\lambda)
  \nonumber \\
  & \times \int_{-\theta_0}^{\theta_0} \! d\theta_{\bk,\bq} \; \frac{m^*}{k\hbar^2} 
  \frac{q_+^e f(q_+^e) + q_-^e f(q_-^e)}{
    \sqrt{\cos^2{\theta_{\bk,\bq}} - \frac{\hbar\omega_\lambda}{\varepsilon_\bk}}}.
\end{align}

\subsection{Collision integral for optical phonon scattering}

In the following two sections, analytic expressions for collision integral in
the case of zero- and first-order optical deformation potential coupling are
given. For the Fr{\"o}hlich interaction the integration over $\theta$ can not
be reduced to a simple analytic form and is therefore evaluated numerically.

\subsubsection{Zero-order deformation potential}

For the zero-order deformation potential coupling in Eq.~\eqref{eq:M_optical},
the $\theta$-integration in the out-scattering part of the collision integral
yields a factor of $2\pi$ resulting in the following form
\begin{align}
  \label{eq:I_out_zero}
  & {I'}_\text{inel}^\text{out} [\phi_k] = 
  - \frac{m^* D_\lambda^2}{2\hbar^2\rho\omega_\lambda}
  \frac{\phi_k}{{1 - f_\bk^0}}
  \nonumber \\
  & \times
  \bigg[
   (1 - f_{\bk + \bq}^0) + (1 - f_{\bk - \bq}^0)
   e^{\hbar\omega_\lambda / k_\text{B}T}
          \Theta(\varepsilon_{\bk} - \hbar\omega_{\lambda} )
   \bigg] N_{\lambda}^0 .
\end{align}
For the in-scattering part, the $q$-independence of the deformation potential
results in a vanishing $\bq$-sum and the in-scattering contribution is zero,
i.e.
\begin{equation}
  \label{eq:I_in_zero}
  {I'}_\text{inel}^\text{in}[\phi_k] = 0  .
\end{equation}

\subsubsection{First-order deformation potential} 

For first-order deformation potentials, we find for the out-scattering part
\begin{align}
  \label{eq:I_out_first}
  {I'}_\text{inel}^\text{out} & [\phi_k] = 
  - \frac{m^{*2} \Xi_\lambda^2}{\hbar^4\rho\omega_\lambda}
  \frac{\phi_k}{{1 - f_\bk^0}} \times 
  \bigg[ 
   (1 - f_{\bk + \bq}^0)( 2\varepsilon_\bk + \hbar\omega_\lambda) \bigg. \nonumber \\
   & + (1 - f_{\bk - \bq}^0) ( 2\varepsilon_\bk - \hbar\omega_\lambda)  
     e^{\hbar\omega_\lambda / k_\text{B}T} 
          \Theta(\varepsilon_{\bk} - \hbar\omega_{\lambda} )
   \bigg] N_{\lambda}^0 .
\end{align}
For the in-scattering part, the result for the first and second term inside the
square brackets in Eq.~\eqref{eq:I_in_1} is
\begin{equation}
  \label{eq:I_in_first1}
  {I'}_\text{inel}^\text{in} [\phi_k]  =  
  - \frac{m^{*2} \Xi_\lambda^2}{\hbar^4\rho\omega_\lambda} 
  \frac{1 - f_{\bk+\bq}^0}{1 - f_\bk^0} 
  \frac{\phi_{k+q}}{\sqrt{1 + \frac{\hbar\omega_\lambda}{\varepsilon_\bk}}}
   N_{\lambda}^0 (\varepsilon_\bk + \hbar\omega_\lambda)
\end{equation}
and
\begin{align}
  \label{eq:I_in_first2}
  {I'}_\text{inel}^\text{in} & [\phi_k]  =  
  - \frac{m^{*2} \Xi_\lambda^2}{\hbar^4\rho\omega_\lambda} 
  \frac{1 - f_{\bk - \bq}^0}{1 - f_\bk^0} 
  \frac{\phi_{k-q}}{\sqrt{1 - \frac{\hbar\omega_\lambda}{\varepsilon_\bk}}}
  \nonumber \\
  & \times
  (1 + N_{\lambda}^0 )
  \left( \varepsilon_\bk - \hbar\omega_\lambda \right) 
  \Theta(\varepsilon_{\bk} - \hbar\omega_{\lambda} ) 
  \nonumber \\
  & \times \frac{1}{\pi}
    \left( \frac{\pi}{2} + 
           \arccos{\sqrt{\frac{\hbar\omega_\lambda}{\varepsilon_\bk}}} +
           \arcsin{\sqrt{\frac{\hbar\omega_\lambda}{\varepsilon_\bk}}} 
    \right) ,
\end{align}
which accounts for absorption and emission, respectively.


\begin{thebibliography}{56}
\expandafter\ifx\csname natexlab\endcsname\relax\def\natexlab#1{#1}\fi
\expandafter\ifx\csname bibnamefont\endcsname\relax
  \def\bibnamefont#1{#1}\fi
\expandafter\ifx\csname bibfnamefont\endcsname\relax
  \def\bibfnamefont#1{#1}\fi
\expandafter\ifx\csname citenamefont\endcsname\relax
  \def\citenamefont#1{#1}\fi
\expandafter\ifx\csname url\endcsname\relax
  \def\url#1{\texttt{#1}}\fi
\expandafter\ifx\csname urlprefix\endcsname\relax\def\urlprefix{URL }\fi
\providecommand{\bibinfo}[2]{#2}
\providecommand{\eprint}[2][]{\url{#2}}

\bibitem[{\citenamefont{Geim and Novoselov}(2007)}]{Geim:Graphene}
\bibinfo{author}{\bibfnamefont{A.~K.} \bibnamefont{Geim}} \bibnamefont{and}
  \bibinfo{author}{\bibfnamefont{K.~S.} \bibnamefont{Novoselov}},
  \bibinfo{journal}{Nature Mat.} \textbf{\bibinfo{volume}{6}},
  \bibinfo{pages}{183} (\bibinfo{year}{2007}).

\bibitem[{\citenamefont{Neto et~al.}(2009)\citenamefont{Neto, Guinea, Peres,
  Novoselov, and Geim}}]{RMP:Graphene}
\bibinfo{author}{\bibfnamefont{A.~H.~C.} \bibnamefont{Neto}},
  \bibinfo{author}{\bibfnamefont{F.}~\bibnamefont{Guinea}},
  \bibinfo{author}{\bibfnamefont{N.~M.~R.} \bibnamefont{Peres}},
  \bibinfo{author}{\bibfnamefont{K.~S.} \bibnamefont{Novoselov}},
  \bibnamefont{and} \bibinfo{author}{\bibfnamefont{A.~K.} \bibnamefont{Geim}},
  \bibinfo{journal}{Rev. Mod. Phys.} \textbf{\bibinfo{volume}{81}},
  \bibinfo{pages}{109} (\bibinfo{year}{2009}).

\bibitem[{\citenamefont{Sarma et~al.}(2011)\citenamefont{Sarma, Adam, Hwang,
  and Rossi}}]{Sarma:RMP}
\bibinfo{author}{\bibfnamefont{S.~D.} \bibnamefont{Sarma}},
  \bibinfo{author}{\bibfnamefont{S.}~\bibnamefont{Adam}},
  \bibinfo{author}{\bibfnamefont{E.~H.} \bibnamefont{Hwang}}, \bibnamefont{and}
  \bibinfo{author}{\bibfnamefont{E.}~\bibnamefont{Rossi}},
  \bibinfo{journal}{Rev. Mod. Phys.} \textbf{\bibinfo{volume}{83}},
  \bibinfo{pages}{407} (\bibinfo{year}{2011}).

\bibitem[{\citenamefont{Neto and Novoselov}(2011)}]{Novoselov:RPP}
\bibinfo{author}{\bibfnamefont{A.~H.~C.} \bibnamefont{Neto}} \bibnamefont{and}
  \bibinfo{author}{\bibfnamefont{K.}~\bibnamefont{Novoselov}},
  \bibinfo{journal}{Rep. Prog. Phys.} \textbf{\bibinfo{volume}{74}},
  \bibinfo{pages}{1} (\bibinfo{year}{2011}).

\bibitem[{\citenamefont{Novoselov et~al.}(2005)\citenamefont{Novoselov, Jiang,
  Schedin, Booth, Khotkevich, Morozov, and Geim}}]{Geim:2D}
\bibinfo{author}{\bibfnamefont{K.~S.} \bibnamefont{Novoselov}},
  \bibinfo{author}{\bibfnamefont{D.}~\bibnamefont{Jiang}},
  \bibinfo{author}{\bibfnamefont{F.}~\bibnamefont{Schedin}},
  \bibinfo{author}{\bibfnamefont{T.~J.} \bibnamefont{Booth}},
  \bibinfo{author}{\bibfnamefont{V.~V.} \bibnamefont{Khotkevich}},
  \bibinfo{author}{\bibfnamefont{S.~V.} \bibnamefont{Morozov}},
  \bibnamefont{and} \bibinfo{author}{\bibfnamefont{A.~K.} \bibnamefont{Geim}},
  \bibinfo{journal}{PNAS} \textbf{\bibinfo{volume}{102}},
  \bibinfo{pages}{10451} (\bibinfo{year}{2005}).

\bibitem[{\citenamefont{Ayari et~al.}(2007)\citenamefont{Ayari, Cobas,
  Ogundadegbe, and Fuhrer}}]{Fuhrer:Ultrathin}
\bibinfo{author}{\bibfnamefont{A.}~\bibnamefont{Ayari}},
  \bibinfo{author}{\bibfnamefont{E.}~\bibnamefont{Cobas}},
  \bibinfo{author}{\bibfnamefont{O.}~\bibnamefont{Ogundadegbe}},
  \bibnamefont{and} \bibinfo{author}{\bibfnamefont{M.~S.}
  \bibnamefont{Fuhrer}}, \bibinfo{journal}{J. Appl. Phys.}
  \textbf{\bibinfo{volume}{101}}, \bibinfo{pages}{014507}
  (\bibinfo{year}{2007}).

\bibitem[{\citenamefont{Matte et~al.}(2010)\citenamefont{Matte, Gomathi, Manna,
  Late, Datta, Pati, and Rao}}]{Pati:Analogues}
\bibinfo{author}{\bibfnamefont{H.~S. S.~R.} \bibnamefont{Matte}},
  \bibinfo{author}{\bibfnamefont{A.}~\bibnamefont{Gomathi}},
  \bibinfo{author}{\bibfnamefont{A.~K.} \bibnamefont{Manna}},
  \bibinfo{author}{\bibfnamefont{D.~J.} \bibnamefont{Late}},
  \bibinfo{author}{\bibfnamefont{R.}~\bibnamefont{Datta}},
  \bibinfo{author}{\bibfnamefont{S.~K.} \bibnamefont{Pati}}, \bibnamefont{and}
  \bibinfo{author}{\bibfnamefont{C.~N.~R.} \bibnamefont{Rao}},
  \bibinfo{journal}{Angew. Chem.} \textbf{\bibinfo{volume}{122}},
  \bibinfo{pages}{4153} (\bibinfo{year}{2010}).

\bibitem[{\citenamefont{Radisavljevic et~al.}(2011)\citenamefont{Radisavljevic,
  Radenovic, Brivio, Giacometti, and Kis}}]{Kis:MoS2Transistor}
\bibinfo{author}{\bibfnamefont{B.}~\bibnamefont{Radisavljevic}},
  \bibinfo{author}{\bibfnamefont{A.}~\bibnamefont{Radenovic}},
  \bibinfo{author}{\bibfnamefont{J.}~\bibnamefont{Brivio}},
  \bibinfo{author}{\bibfnamefont{V.}~\bibnamefont{Giacometti}},
  \bibnamefont{and} \bibinfo{author}{\bibfnamefont{A.}~\bibnamefont{Kis}},
  \bibinfo{journal}{Nature Nano.} \textbf{\bibinfo{volume}{6}},
  \bibinfo{pages}{147} (\bibinfo{year}{2011}).

\bibitem[{\citenamefont{Lee et~al.}(2010)\citenamefont{Lee, Yan, Brus, Heinz,
  Hone, and Ryu}}]{Ryu:Anomalous}
\bibinfo{author}{\bibfnamefont{C.}~\bibnamefont{Lee}},
  \bibinfo{author}{\bibfnamefont{H.}~\bibnamefont{Yan}},
  \bibinfo{author}{\bibfnamefont{L.~E.} \bibnamefont{Brus}},
  \bibinfo{author}{\bibfnamefont{T.~F.} \bibnamefont{Heinz}},
  \bibinfo{author}{\bibfnamefont{J.}~\bibnamefont{Hone}}, \bibnamefont{and}
  \bibinfo{author}{\bibfnamefont{S.}~\bibnamefont{Ryu}}, \bibinfo{journal}{ACS
  Nano} \textbf{\bibinfo{volume}{4}}, \bibinfo{pages}{2695}
  (\bibinfo{year}{2010}).

\bibitem[{\citenamefont{Mak et~al.}(2010)\citenamefont{Mak, Lee, Hone, Shan,
  and Heinz}}]{Heinz:ThinMoS2}
\bibinfo{author}{\bibfnamefont{K.~F.} \bibnamefont{Mak}},
  \bibinfo{author}{\bibfnamefont{C.}~\bibnamefont{Lee}},
  \bibinfo{author}{\bibfnamefont{J.}~\bibnamefont{Hone}},
  \bibinfo{author}{\bibfnamefont{J.}~\bibnamefont{Shan}}, \bibnamefont{and}
  \bibinfo{author}{\bibfnamefont{T.~F.} \bibnamefont{Heinz}},
  \bibinfo{journal}{Phys. Rev. Lett.} \textbf{\bibinfo{volume}{105}},
  \bibinfo{pages}{136805} (\bibinfo{year}{2010}).

\bibitem[{\citenamefont{Korn et~al.}(2011)\citenamefont{Korn, Heydrich, Hirmer,
  Schmutzler, and Sch{\"u}ller}}]{Schuller:Photocarrier}
\bibinfo{author}{\bibfnamefont{T.}~\bibnamefont{Korn}},
  \bibinfo{author}{\bibfnamefont{S.}~\bibnamefont{Heydrich}},
  \bibinfo{author}{\bibfnamefont{M.}~\bibnamefont{Hirmer}},
  \bibinfo{author}{\bibfnamefont{J.}~\bibnamefont{Schmutzler}},
  \bibnamefont{and}
  \bibinfo{author}{\bibfnamefont{C.}~\bibnamefont{Sch{\"u}ller}},
  \bibinfo{journal}{Appl. Phys. Lett.} \textbf{\bibinfo{volume}{99}},
  \bibinfo{pages}{102109} (\bibinfo{year}{2011}).

\bibitem[{\citenamefont{Splendiani et~al.}(2010)\citenamefont{Splendiani, Sun,
  Zhang, Li, Kim, Chim, Galli, and Wang}}]{Wang:Emerging}
\bibinfo{author}{\bibfnamefont{A.}~\bibnamefont{Splendiani}},
  \bibinfo{author}{\bibfnamefont{L.}~\bibnamefont{Sun}},
  \bibinfo{author}{\bibfnamefont{Y.}~\bibnamefont{Zhang}},
  \bibinfo{author}{\bibfnamefont{T.}~\bibnamefont{Li}},
  \bibinfo{author}{\bibfnamefont{J.}~\bibnamefont{Kim}},
  \bibinfo{author}{\bibfnamefont{C.-Y.} \bibnamefont{Chim}},
  \bibinfo{author}{\bibfnamefont{G.}~\bibnamefont{Galli}}, \bibnamefont{and}
  \bibinfo{author}{\bibfnamefont{F.}~\bibnamefont{Wang}},
  \bibinfo{journal}{Nano. Lett.} \textbf{\bibinfo{volume}{10}},
  \bibinfo{pages}{1271} (\bibinfo{year}{2010}).

\bibitem[{\citenamefont{Yoon et~al.}(2011)\citenamefont{Yoon, Ganapathi, and
  Salahuddin}}]{Salahuddin:HowGood}
\bibinfo{author}{\bibfnamefont{Y.}~\bibnamefont{Yoon}},
  \bibinfo{author}{\bibfnamefont{K.}~\bibnamefont{Ganapathi}},
  \bibnamefont{and}
  \bibinfo{author}{\bibfnamefont{S.}~\bibnamefont{Salahuddin}},
  \bibinfo{journal}{Nano. Lett.} \textbf{\bibinfo{volume}{11}},
  \bibinfo{pages}{3768} (\bibinfo{year}{2011}).

\bibitem[{\citenamefont{Fivaz and Mooser}(1967)}]{Mooser:Mobility}
\bibinfo{author}{\bibfnamefont{R.}~\bibnamefont{Fivaz}} \bibnamefont{and}
  \bibinfo{author}{\bibfnamefont{E.}~\bibnamefont{Mooser}},
  \bibinfo{journal}{Phys. Rev.} \textbf{\bibinfo{volume}{163}},
  \bibinfo{pages}{743} (\bibinfo{year}{1967}).

\bibitem[{\citenamefont{Hwang and Sarma}(2008)}]{Sarma:100}
\bibinfo{author}{\bibfnamefont{E.~H.} \bibnamefont{Hwang}} \bibnamefont{and}
  \bibinfo{author}{\bibfnamefont{S.} \bibnamefont{Das~Sarma}},
  \bibinfo{journal}{Phys. Rev. B} \textbf{\bibinfo{volume}{77}},
  \bibinfo{pages}{235437} (\bibinfo{year}{2008}).

\bibitem[{\citenamefont{Kawamura and Sarma}(1990)}]{Sarma:Hetero1}
\bibinfo{author}{\bibfnamefont{T.}~\bibnamefont{Kawamura}} \bibnamefont{and}
  \bibinfo{author}{\bibfnamefont{S.} \bibnamefont{Das~Sarma}},
  \bibinfo{journal}{Phys. Rev. B} \textbf{\bibinfo{volume}{42}},
  \bibinfo{pages}{3725} (\bibinfo{year}{1990}).

\bibitem[{\citenamefont{Hwang et~al.}(2007)\citenamefont{Hwang, Adam, and
  Sarma}}]{Sarma:2DGraphene}
\bibinfo{author}{\bibfnamefont{E.~H.} \bibnamefont{Hwang}},
  \bibinfo{author}{\bibfnamefont{S.}~\bibnamefont{Adam}}, \bibnamefont{and}
  \bibinfo{author}{\bibfnamefont{S.} \bibnamefont{Das~Sarma}},
  \bibinfo{journal}{Phys. Rev. Lett.} \textbf{\bibinfo{volume}{98}},
  \bibinfo{pages}{186806} (\bibinfo{year}{2007}).

\bibitem[{\citenamefont{Nomura and MacDonald}(2007)}]{MacDonald:Massless}
\bibinfo{author}{\bibfnamefont{K.}~\bibnamefont{Nomura}} \bibnamefont{and}
  \bibinfo{author}{\bibfnamefont{A.~H.} \bibnamefont{MacDonald}},
  \bibinfo{journal}{Phys. Rev. Lett.} \textbf{\bibinfo{volume}{98}},
  \bibinfo{pages}{076602} (\bibinfo{year}{2007}).

\bibitem[{\citenamefont{Chen et~al.}(2008)\citenamefont{Chen, Jang, Xiao,
  Ishigami, and Fuhrer}}]{Fuhrer:GrapheneSiO2}
\bibinfo{author}{\bibfnamefont{J.-H.} \bibnamefont{Chen}},
  \bibinfo{author}{\bibfnamefont{C.}~\bibnamefont{Jang}},
  \bibinfo{author}{\bibfnamefont{S.}~\bibnamefont{Xiao}},
  \bibinfo{author}{\bibfnamefont{M.}~\bibnamefont{Ishigami}}, \bibnamefont{and}
  \bibinfo{author}{\bibfnamefont{M.~S.} \bibnamefont{Fuhrer}},
  \bibinfo{journal}{Nature Nano.} \textbf{\bibinfo{volume}{3}},
  \bibinfo{pages}{206} (\bibinfo{year}{2008}).

\bibitem[{\citenamefont{Konar et~al.}(2010)\citenamefont{Konar, Fang, and
  Jena}}]{Jena:HighKappa}
\bibinfo{author}{\bibfnamefont{A.}~\bibnamefont{Konar}},
  \bibinfo{author}{\bibfnamefont{T.}~\bibnamefont{Fang}}, \bibnamefont{and}
  \bibinfo{author}{\bibfnamefont{D.}~\bibnamefont{Jena}},
  \bibinfo{journal}{Phys. Rev. B} \textbf{\bibinfo{volume}{82}},
  \bibinfo{pages}{115452} (\bibinfo{year}{2010}).

\bibitem[{\citenamefont{Li et~al.}(2010)\citenamefont{Li, Barry, Zavada,
  Nardelli, and Kim}}]{Kim:SO}
\bibinfo{author}{\bibfnamefont{X.}~\bibnamefont{Li}},
  \bibinfo{author}{\bibfnamefont{E.~A.} \bibnamefont{Barry}},
  \bibinfo{author}{\bibfnamefont{J.~M.} \bibnamefont{Zavada}},
  \bibinfo{author}{\bibfnamefont{M.~B.} \bibnamefont{Nardelli}},
  \bibnamefont{and} \bibinfo{author}{\bibfnamefont{K.~W.} \bibnamefont{Kim}},
  \bibinfo{journal}{Appl. Phys. Lett.} \textbf{\bibinfo{volume}{97}},
  \bibinfo{pages}{232405} (\bibinfo{year}{2010}).

\bibitem[{\citenamefont{Heo et~al.}(2011)\citenamefont{Heo, Chung, Lee, Yang,
  Seo, Shin, Chung, Seo, Hwang, and Sarma}}]{Sarma:Nonmonotonic}
\bibinfo{author}{\bibfnamefont{J.}~\bibnamefont{Heo}},
  \bibinfo{author}{\bibfnamefont{H.~J.} \bibnamefont{Chung}},
  \bibinfo{author}{\bibfnamefont{S.-H.} \bibnamefont{Lee}},
  \bibinfo{author}{\bibfnamefont{H.}~\bibnamefont{Yang}},
  \bibinfo{author}{\bibfnamefont{D.~H.} \bibnamefont{Seo}},
  \bibinfo{author}{\bibfnamefont{J.~K.} \bibnamefont{Shin}},
  \bibinfo{author}{\bibfnamefont{U.-I.} \bibnamefont{Chung}},
  \bibinfo{author}{\bibfnamefont{S.}~\bibnamefont{Seo}},
  \bibinfo{author}{\bibfnamefont{E.~H.} \bibnamefont{Hwang}}, \bibnamefont{and}
  \bibinfo{author}{\bibfnamefont{S.} \bibnamefont{Das~Sarma}},
  \bibinfo{journal}{Phys. Rev. B} \textbf{\bibinfo{volume}{84}},
  \bibinfo{pages}{035421} (\bibinfo{year}{2011}).

\bibitem[{\citenamefont{Fivaz}(1969)}]{Fivaz:Dimensionality}
\bibinfo{author}{\bibfnamefont{R.~C.} \bibnamefont{Fivaz}},
  \bibinfo{journal}{NUOVO CIMENTO} \textbf{\bibinfo{volume}{63 B}},
  \bibinfo{pages}{10} (\bibinfo{year}{1969}).

\bibitem[{\citenamefont{Schmid}(1974)}]{Schmid:Layered}
\bibinfo{author}{\bibfnamefont{P.}~\bibnamefont{Schmid}},
  \bibinfo{journal}{NUOVO CIMENTO} \textbf{\bibinfo{volume}{21 B}},
  \bibinfo{pages}{258} (\bibinfo{year}{1974}).

\bibitem[{\citenamefont{Mortensen et~al.}(2005)\citenamefont{Mortensen, Hansen,
  and Jacobsen}}]{GPAW}
\bibinfo{author}{\bibfnamefont{J.~J.} \bibnamefont{Mortensen}},
  \bibinfo{author}{\bibfnamefont{L.~B.} \bibnamefont{Hansen}},
  \bibnamefont{and} \bibinfo{author}{\bibfnamefont{K.~W.}
  \bibnamefont{Jacobsen}}, \bibinfo{journal}{Phys. Rev. B}
  \textbf{\bibinfo{volume}{71}}, \bibinfo{pages}{035109}
  (\bibinfo{year}{2005}).

\bibitem[{\citenamefont{Larsen et~al.}(2009)\citenamefont{Larsen, Vanin,
  Mortensen, Thygesen, and Jacobsen}}]{GPAW1}
\bibinfo{author}{\bibfnamefont{A.~H.} \bibnamefont{Larsen}},
  \bibinfo{author}{\bibfnamefont{M.}~\bibnamefont{Vanin}},
  \bibinfo{author}{\bibfnamefont{J.~J.} \bibnamefont{Mortensen}},
  \bibinfo{author}{\bibfnamefont{K.~S.} \bibnamefont{Thygesen}},
  \bibnamefont{and} \bibinfo{author}{\bibfnamefont{K.~W.}
  \bibnamefont{Jacobsen}}, \bibinfo{journal}{Phys. Rev. B}
  \textbf{\bibinfo{volume}{80}}, \bibinfo{pages}{195112}
  (\bibinfo{year}{2009}).

\bibitem[{\citenamefont{Enkovaara et~al.}(2010)\citenamefont{Enkovaara,
  Rostgaard, Mortensen, Chen, Dulak, Ferrighi, Gavnholt, Glinsvad, Haikola,
  Hansen et~al.}}]{GPAW2}
\bibinfo{author}{\bibfnamefont{J.~.} \bibnamefont{Enkovaara}},
  \bibinfo{author}{\bibfnamefont{C.}~\bibnamefont{Rostgaard}},
  \bibinfo{author}{\bibfnamefont{J.~J.} \bibnamefont{Mortensen}},
  \bibinfo{author}{\bibfnamefont{J.}~\bibnamefont{Chen}},
  \bibinfo{author}{\bibfnamefont{M.}~\bibnamefont{Dulak}},
  \bibinfo{author}{\bibfnamefont{L.}~\bibnamefont{Ferrighi}},
  \bibinfo{author}{\bibfnamefont{J.}~\bibnamefont{Gavnholt}},
  \bibinfo{author}{\bibfnamefont{C.}~\bibnamefont{Glinsvad}},
  \bibinfo{author}{\bibfnamefont{V.}~\bibnamefont{Haikola}},
  \bibinfo{author}{\bibfnamefont{H.~A.} \bibnamefont{Hansen}},
  \bibnamefont{et~al.}, \bibinfo{journal}{J. Phys.: Condens. Matter}
  \textbf{\bibinfo{volume}{22}}, \bibinfo{pages}{253202}
  (\bibinfo{year}{2010}).

\bibitem[{\citenamefont{Leb{\`e}gue and Eriksson}(2009)}]{Eriksson:2D}
\bibinfo{author}{\bibfnamefont{S.}~\bibnamefont{Leb{\`e}gue}} \bibnamefont{and}
  \bibinfo{author}{\bibfnamefont{O.}~\bibnamefont{Eriksson}},
  \bibinfo{journal}{Phys. Rev. B} \textbf{\bibinfo{volume}{79}},
  \bibinfo{pages}{115409} (\bibinfo{year}{2009}).

\bibitem[{\citenamefont{Han et~al.}(2011)\citenamefont{Han, Kwon, Kim, Ryu,
  Yun, Kim, Hwang, Kang, Baik, Shin et~al.}}]{Hong:Interlayer}
\bibinfo{author}{\bibfnamefont{S.~W.} \bibnamefont{Han}},
  \bibinfo{author}{\bibfnamefont{H.}~\bibnamefont{Kwon}},
  \bibinfo{author}{\bibfnamefont{S.~K.} \bibnamefont{Kim}},
  \bibinfo{author}{\bibfnamefont{S.}~\bibnamefont{Ryu}},
  \bibinfo{author}{\bibfnamefont{W.~S.} \bibnamefont{Yun}},
  \bibinfo{author}{\bibfnamefont{D.~H.} \bibnamefont{Kim}},
  \bibinfo{author}{\bibfnamefont{J.~H.} \bibnamefont{Hwang}},
  \bibinfo{author}{\bibfnamefont{J.-S.} \bibnamefont{Kang}},
  \bibinfo{author}{\bibfnamefont{J.}~\bibnamefont{Baik}},
  \bibinfo{author}{\bibfnamefont{H.~J.} \bibnamefont{Shin}},
  \bibnamefont{et~al.}, \bibinfo{journal}{Phys. Rev. B}
  \textbf{\bibinfo{volume}{84}}, \bibinfo{pages}{045409}
  (\bibinfo{year}{2011}).

\bibitem[{\citenamefont{Alf{\'e}}(2009)}]{Alfe:SDM}
\bibinfo{author}{\bibfnamefont{D.}~\bibnamefont{Alf{\'e}}},
  \bibinfo{journal}{Comput. Phys. Commun.} \textbf{\bibinfo{volume}{180}},
  \bibinfo{pages}{2622} (\bibinfo{year}{2009}).

\bibitem[{\citenamefont{Ataca et~al.}(2011)\citenamefont{Ataca, Topsakal,
  Akt{\"u}rk, and Ciraci}}]{Ciraci:Lattice}
\bibinfo{author}{\bibfnamefont{C.}~\bibnamefont{Ataca}},
  \bibinfo{author}{\bibfnamefont{M.}~\bibnamefont{Topsakal}},
  \bibinfo{author}{\bibfnamefont{E.}~\bibnamefont{Akt{\"u}rk}},
  \bibnamefont{and} \bibinfo{author}{\bibfnamefont{S.}~\bibnamefont{Ciraci}},
  \bibinfo{journal}{J. Phys. Chem. C} \textbf{\bibinfo{volume}{115}},
  \bibinfo{pages}{16354} (\bibinfo{year}{2011}).

\bibitem[{\citenamefont{Giannozzi et~al.}(1991)\citenamefont{Giannozzi,
  de~Gironcoli, Pavone, and Baroni}}]{Baroni:PhononSemi}
\bibinfo{author}{\bibfnamefont{P.}~\bibnamefont{Giannozzi}},
  \bibinfo{author}{\bibfnamefont{S.}~\bibnamefont{de~Gironcoli}},
  \bibinfo{author}{\bibfnamefont{P.}~\bibnamefont{Pavone}}, \bibnamefont{and}
  \bibinfo{author}{\bibfnamefont{S.}~\bibnamefont{Baroni}},
  \bibinfo{journal}{Phys. Rev. B} \textbf{\bibinfo{volume}{43}},
  \bibinfo{pages}{7231} (\bibinfo{year}{1991}).

\bibitem[{\citenamefont{Gonze and Lee}(1997)}]{Gonze:DynamicalMatrices}
\bibinfo{author}{\bibfnamefont{X.}~\bibnamefont{Gonze}} \bibnamefont{and}
  \bibinfo{author}{\bibfnamefont{C.}~\bibnamefont{Lee}},
  \bibinfo{journal}{Phys. Rev. B} \textbf{\bibinfo{volume}{55}},
  \bibinfo{pages}{10355} (\bibinfo{year}{1997}).

\bibitem[{\citenamefont{Wang et~al.}(2010)\citenamefont{Wang, Wang, Wang, Mei,
  Shang, Chen, and Liu}}]{Wang:MixedSpace}
\bibinfo{author}{\bibfnamefont{Y.}~\bibnamefont{Wang}},
  \bibinfo{author}{\bibfnamefont{J.~J.} \bibnamefont{Wang}},
  \bibinfo{author}{\bibfnamefont{W.~Y.} \bibnamefont{Wang}},
  \bibinfo{author}{\bibfnamefont{Z.~G.} \bibnamefont{Mei}},
  \bibinfo{author}{\bibfnamefont{S.~L.} \bibnamefont{Shang}},
  \bibinfo{author}{\bibfnamefont{L.~Q.} \bibnamefont{Chen}}, \bibnamefont{and}
  \bibinfo{author}{\bibfnamefont{Z.~K.} \bibnamefont{Liu}},
  \bibinfo{journal}{J. Phys.: Condens. Matter} \textbf{\bibinfo{volume}{22}},
  \bibinfo{pages}{202201} (\bibinfo{year}{2010}).

\bibitem[{\citenamefont{S{\'a}nchez-Portal and
  Hern{\'a}ndez}(2002)}]{Hernandez:MonolayerBN}
\bibinfo{author}{\bibfnamefont{D.}~\bibnamefont{S{\'a}nchez-Portal}}
  \bibnamefont{and}
  \bibinfo{author}{\bibfnamefont{E.}~\bibnamefont{Hern{\'a}ndez}},
  \bibinfo{journal}{Phys. Rev. B} \textbf{\bibinfo{volume}{66}},
  \bibinfo{pages}{235415} (\bibinfo{year}{2002}).

\bibitem[{foo({\natexlab{a}})}]{footnote1}
\bibinfo{note}{The calculation of the electron-phonon interaction has been
  performed using a $9\times 9$ supercell and a DZP basis for the electronic
  Bloch states. For the polar LO phonon, a $17 \times 17$ supercell was
  required to capture the long-wavelength limiting behavior of the
  electron-phonon coupling due to long-range Coulomb interactions. This
  calculation must be done with Dirichlet boundary conditions, i.e. $V(\br) =
  0$, on the boundaries in the non-periodic direction perpendicular to the
  layer in order to avoid interlayer contributions to the potential in the
  long-wavelength limit.}

\bibitem[{\citenamefont{Kawamura and Sarma}(1992)}]{Sarma:Hetero2}
\bibinfo{author}{\bibfnamefont{T.}~\bibnamefont{Kawamura}} \bibnamefont{and}
  \bibinfo{author}{\bibfnamefont{S.} \bibnamefont{Das~Sarma}},
  \bibinfo{journal}{Phys. Rev. B} \textbf{\bibinfo{volume}{45}},
  \bibinfo{pages}{3612} (\bibinfo{year}{1992}).

\bibitem[{\citenamefont{Ferry}(2000)}]{Ferry}
\bibinfo{author}{\bibfnamefont{D.~K.} \bibnamefont{Ferry}},
  \emph{\bibinfo{title}{Semiconductor Transport}} (\bibinfo{publisher}{Taylor
  and Francis}, \bibinfo{address}{New York}, \bibinfo{year}{2000}).

\bibitem[{foo({\natexlab{b}})}]{footnote2}
\bibinfo{note}{For non-degenerate carriers at room temperature
  $\expect{\varepsilon_\bk}\sim 26$ meV implying that $k\sim 0.03\times 2\pi/a$
  is a good measure for the wave vector of the carriers. With a maximum phonon
  frequency of $\sim$50~meV, the phonon wave vectors will be restricted to the
  interval $0 \le q \le q_\text{max}$ where $q_\text{max} \simeq 0.1 \times
  2\pi/a$ is the phonon wave vector in a backscattering process with $q=2k$.}

\bibitem[{\citenamefont{Madelung}(1996)}]{Madelung}
\bibinfo{author}{\bibfnamefont{O.}~\bibnamefont{Madelung}},
  \emph{\bibinfo{title}{Introduction to Solid State Physics}}
  (\bibinfo{publisher}{Springer}, \bibinfo{address}{Berlin},
  \bibinfo{year}{1996}).

\bibitem[{\citenamefont{Sjakste et~al.}(2006)\citenamefont{Sjakste, Tyuterev,
  and Vast}}]{Sjakste:GaAsInter}
\bibinfo{author}{\bibfnamefont{J.}~\bibnamefont{Sjakste}},
  \bibinfo{author}{\bibfnamefont{V.}~\bibnamefont{Tyuterev}}, \bibnamefont{and}
  \bibinfo{author}{\bibfnamefont{N.}~\bibnamefont{Vast}},
  \bibinfo{journal}{Phys. Rev. B} \textbf{\bibinfo{volume}{74}},
  \bibinfo{pages}{235216} (\bibinfo{year}{2006}).

\bibitem[{\citenamefont{Tyuterev et~al.}(2010)\citenamefont{Tyuterev, Sjakste,
  and Vast}}]{Sjakste:Silicon}
\bibinfo{author}{\bibfnamefont{V.}~\bibnamefont{Tyuterev}},
  \bibinfo{author}{\bibfnamefont{J.}~\bibnamefont{Sjakste}}, \bibnamefont{and}
  \bibinfo{author}{\bibfnamefont{N.}~\bibnamefont{Vast}},
  \bibinfo{journal}{Phys. Rev. B} \textbf{\bibinfo{volume}{81}},
  \bibinfo{pages}{245212} (\bibinfo{year}{2010}).

\bibitem[{\citenamefont{Borysenko et~al.}(2010)\citenamefont{Borysenko, Mullen,
  Barry, Paul, Semenov, Zavada, Nardelli, and Kim}}]{Kim:ElPhGraphene}
\bibinfo{author}{\bibfnamefont{K.~M.} \bibnamefont{Borysenko}},
  \bibinfo{author}{\bibfnamefont{J.~T.} \bibnamefont{Mullen}},
  \bibinfo{author}{\bibfnamefont{E.~A.} \bibnamefont{Barry}},
  \bibinfo{author}{\bibfnamefont{S.}~\bibnamefont{Paul}},
  \bibinfo{author}{\bibfnamefont{Y.~G.} \bibnamefont{Semenov}},
  \bibinfo{author}{\bibfnamefont{J.~M.} \bibnamefont{Zavada}},
  \bibinfo{author}{\bibfnamefont{M.~B.} \bibnamefont{Nardelli}},
  \bibnamefont{and} \bibinfo{author}{\bibfnamefont{K.~W.} \bibnamefont{Kim}},
  \bibinfo{journal}{Phys. Rev. B} \textbf{\bibinfo{volume}{81}},
  \bibinfo{pages}{121412} (\bibinfo{year}{2010}).

\bibitem[{\citenamefont{Mori and Ando}(1989)}]{Ando:ElphHetero}
\bibinfo{author}{\bibfnamefont{N.}~\bibnamefont{Mori}} \bibnamefont{and}
  \bibinfo{author}{\bibfnamefont{T.}~\bibnamefont{Ando}},
  \bibinfo{journal}{Phys. Rev. B} \textbf{\bibinfo{volume}{40}},
  \bibinfo{pages}{6175} (\bibinfo{year}{1989}).

\bibitem[{\citenamefont{R{\"u}cker et~al.}(1992)\citenamefont{R{\"u}cker,
  Molinari, and Lugli}}]{Lugli:Microscopic}
\bibinfo{author}{\bibfnamefont{H.}~\bibnamefont{R{\"u}cker}},
  \bibinfo{author}{\bibfnamefont{E.}~\bibnamefont{Molinari}}, \bibnamefont{and}
  \bibinfo{author}{\bibfnamefont{P.}~\bibnamefont{Lugli}},
  \bibinfo{journal}{Phys. Rev. B} \textbf{\bibinfo{volume}{45}},
  \bibinfo{pages}{6747} (\bibinfo{year}{1992}).

\bibitem[{\citenamefont{Smith and Jensen}(1989)}]{SmithJensen}
\bibinfo{author}{\bibfnamefont{H.}~\bibnamefont{Smith}} \bibnamefont{and}
  \bibinfo{author}{\bibfnamefont{H.~H.} \bibnamefont{Jensen}},
  \emph{\bibinfo{title}{Transport Phenomena}} (\bibinfo{publisher}{Oxford},
  \bibinfo{year}{1989}).

\bibitem[{\citenamefont{Rode}(1975)}]{Rode}
\bibinfo{author}{\bibfnamefont{D.~L.} \bibnamefont{Rode}}, in
  \emph{\bibinfo{booktitle}{Semiconductors and Semimetals}}, edited by
  \bibinfo{editor}{\bibfnamefont{R.~K.} \bibnamefont{Willardson}}
  \bibnamefont{and} \bibinfo{editor}{\bibfnamefont{A.~C.} \bibnamefont{Beer}}
  (\bibinfo{publisher}{Academic Press}, \bibinfo{address}{New York},
  \bibinfo{year}{1975}), vol.~\bibinfo{volume}{10}, pp. \bibinfo{pages}{1--89}.

\bibitem[{\citenamefont{Ferry and Goodnick}(2009)}]{FerryGoodnick}
\bibinfo{author}{\bibfnamefont{D.~K.} \bibnamefont{Ferry}} \bibnamefont{and}
  \bibinfo{author}{\bibfnamefont{S.~M.} \bibnamefont{Goodnick}},
  \emph{\bibinfo{title}{Transport in Nanostructures}}
  (\bibinfo{publisher}{Cambridge University Press},
  \bibinfo{address}{Cambridge}, \bibinfo{year}{2009}), \bibinfo{edition}{2nd}
  ed.

\bibitem[{foo({\natexlab{c}})}]{footnote3}
\bibinfo{note}{In order to define a momentum relaxation time $\tau_\bk$, the
  collision integral in Eq.~\eqref{eq:collision_linearized} must be put on the
  form $\partial_t f_\bk\vert_\text{coll} = -\delta f_\bk/\tau_\bk$. For
  inelastic scattering processes, this is general not possible since the
  in-scattering part of the collision integral depends on the deviation
  function $\psi_{\bk \pm \bq}=\psi(\varepsilon_\bk \pm \hbar\omega_\bq)$.}

\bibitem[{foo({\natexlab{d}})}]{footnote4}
\bibinfo{note}{The sum over $\bq$ in Eqs.~\eqref{eq:tau_elastic}
  and~\eqref{eq:tau_optical} have been evaluated using a lorentzian of width
  $\Gamma=3$ meV to represent the $\delta$-function: $\delta(\varepsilon_{\bk'}
  - \varepsilon_{\bk}) \rightarrow 1 / \pi \times \frac{1}{2}\Gamma / \left[
  (\varepsilon_{\bk'} - \varepsilon_{\bk})^2 + (\frac{1}{2}\Gamma)^2 \right]$.}

\bibitem[{\citenamefont{Jena and Konar}(2007)}]{Konar:Engineering}
\bibinfo{author}{\bibfnamefont{D.}~\bibnamefont{Jena}} \bibnamefont{and}
  \bibinfo{author}{\bibfnamefont{A.}~\bibnamefont{Konar}},
  \bibinfo{journal}{Phys. Rev. Lett.} \textbf{\bibinfo{volume}{98}},
  \bibinfo{pages}{136805} (\bibinfo{year}{2007}).

\bibitem[{\citenamefont{Aulbur et~al.}(2000)\citenamefont{Aulbur, J{\"o}nsson,
  and Wilkins}}]{GW}
\bibinfo{author}{\bibfnamefont{W.~G.} \bibnamefont{Aulbur}},
  \bibinfo{author}{\bibfnamefont{L.}~\bibnamefont{J{\"o}nsson}},
  \bibnamefont{and} \bibinfo{author}{\bibfnamefont{J.}~\bibnamefont{Wilkins}},
  in \emph{\bibinfo{booktitle}{Solid State Physics}}, edited by
  \bibinfo{editor}{\bibfnamefont{F.}~\bibnamefont{Seitz}},
  \bibinfo{editor}{\bibfnamefont{D.}~\bibnamefont{Turnbull}}, \bibnamefont{and}
  \bibinfo{editor}{\bibfnamefont{H.}~\bibnamefont{Ehrenreich}}
  (\bibinfo{publisher}{Academic Press}, \bibinfo{address}{New York},
  \bibinfo{year}{2000}), vol.~\bibinfo{volume}{54}, p.~\bibinfo{pages}{1}.

\bibitem[{\citenamefont{Ataca and Ciraci}(2011)}]{Ciraci:Functionalization}
\bibinfo{author}{\bibfnamefont{C.}~\bibnamefont{Ataca}} \bibnamefont{and}
  \bibinfo{author}{\bibfnamefont{S.}~\bibnamefont{Ciraci}},
  \bibinfo{journal}{J. Phys. Chem. C} \textbf{\bibinfo{volume}{115}},
  \bibinfo{pages}{13303} (\bibinfo{year}{2011}).

\bibitem[{\citenamefont{Olsen et~al.}(2011)\citenamefont{Olsen, Jacobsen, and
  Thygesen}}]{Olsen:MoS2}
\bibinfo{author}{\bibfnamefont{T.}~\bibnamefont{Olsen}},
  \bibinfo{author}{\bibfnamefont{K.~W.} \bibnamefont{Jacobsen}},
  \bibnamefont{and} \bibinfo{author}{\bibfnamefont{K.~S.}
  \bibnamefont{Thygesen}}, \bibinfo{journal}{arXiv:1107.0600v1}
  (\bibinfo{year}{2011}).

\bibitem[{\citenamefont{Giustino et~al.}(2007)\citenamefont{Giustino, Cohen,
  and Louie}}]{Louie:e-ph}
\bibinfo{author}{\bibfnamefont{F.}~\bibnamefont{Giustino}},
  \bibinfo{author}{\bibfnamefont{M.~L.} \bibnamefont{Cohen}}, \bibnamefont{and}
  \bibinfo{author}{\bibfnamefont{S.~G.} \bibnamefont{Louie}},
  \bibinfo{journal}{Phys. Rev. B} \textbf{\bibinfo{volume}{76}},
  \bibinfo{pages}{165108} (\bibinfo{year}{2007}).

\bibitem[{\citenamefont{Bl{\"o}chl}(1994)}]{Blochl:PAW}
\bibinfo{author}{\bibfnamefont{P.~E.} \bibnamefont{Bl{\"o}chl}},
  \bibinfo{journal}{prb} \textbf{\bibinfo{volume}{50}}, \bibinfo{pages}{17953}
  (\bibinfo{year}{1994}).

\end{thebibliography}
\end{document}